\documentclass[twocolumn,eqsecnum,pre,showpacs,floatfix]{revtex4-1}
\usepackage[latin1]{inputenc}
\usepackage[english]{babel}
\usepackage{graphicx}
\usepackage{amsmath}
\usepackage{bm}
\usepackage{amsfonts}
\usepackage{amssymb}
\usepackage[usenames]{color}

\begin{document}

\newcommand{\brm}[1]{\bm{{\rm #1}}}
\newcommand{\Ochange}[1]{{\color{red}{#1}}}
\newcommand{\Ocomment}[1]{{\color{green}{#1}}}
\newcommand{\Oremove}[1]{{\color{yellow}{#1}}}
\newcommand{\Hcomment}[1]{{\color{blue}{#1}}}
\newcommand{\Hchange}[1]{{\color{magenta}{#1}}}

\title{Scaling exponents  for a monkey on a tree\\
-- fractal dimensions of randomly branched polymers}
\author{Hans-Karl Janssen}
\affiliation{Institut f\"ur Theoretische Physik III,
Heinrich-Heine-Universit\"at, 40225 D\"usseldorf, Germany}
\author{Olaf Stenull}
\affiliation{Department of Physics and Astronomy, University of Pennsylvania,
Philadelphia PA 19104, USA}
\date{\today}

\begin{abstract}
We study asymptotic properties of diffusion and other transport processes
(including self-avoiding walks and electrical conduction) on large randomly
branched polymers using renormalized dynamical field theory. We focus on the
swollen phase and the collapse transition, where loops in the polymers are
irrelevant. Here the asymptotic statistics of the polymers is that of lattice
trees, and diffusion on them is reminiscent of the climbing of a monkey on a
tree. We calculate a set of universal scaling exponents including the diffusion
exponent and the fractal dimension of the minimal path to 2-loop order and,
where available, compare them to numerical results.
\end{abstract}

\maketitle

\section{Introduction}

In 1976, de Gennes coined the nickname {\em ant in the labyrinth} for the
problem of random walks on a fractal structure such as a percolation cluster
near the percolation point~\cite{dGen76}. In 1982, Wilke, Gefen, Ilkovic,
Aharony, and Stauffer~\cite{WiGeIlAhSt84} introduced the {\em parasite} problem
as a variant of the former problem in which a random walk takes place on a
lattice animal, i.e., a large percolation cluster in the region right below the
percolation point. The statistics of lattice animals belong to the same
universality classes as the statistics of randomly branched polymers (RBPs) in
dilute solutions in their swollen phase and at the collapse transition
($\Theta$-line), respectively~\cite{LuIs78,PaSo81,JaSt10}. Because cycles or
loops of the animals and RBPs are irrelevant for these universality classes,
they are the same as for lattice trees. In other words, the animal or plant on
which a random walk takes place in these universality classes is tree-like.
Thus, the random walks we are studying in this paper remind us more of a {\em
monkey on a tree} than a {\em parasite} on a loop-containing animal, and we
prefer to glean our nickname from primatology rather than parasitology.

Topologically, trees are $1$-dimensional structures. Hence, the backbone
between two separated points on a tree consists of an single unique self
avoiding walk (SAW). Therefore, all the scaling dimensions $d_{\alpha}$ of the
backbone fractals -- the backbone itself, the minimal, the mean, and the
maximal path, the electrical resistance, the red bonds, \textit{etc}.\ -- are
identical:
\begin{equation}
d_{B}=d_{\min}=d_{\text{SAW}}=d_{\max}=d_{R}=d_{red}\,. \label{DimBB}
\end{equation}
A clever monkey will climb the minimal path with fractal dimension $d_{\min}$
to get a coconut at the end of the path. This is in contrast to diffusion,
which corresponds to an erratic motion of a monkey. Diffusion, on a fractal
medium with fractal dimension $d_{f}$ is described by the scaling law
\begin{equation}
\langle\bigl(\mathbf{r}(t)-\mathbf{r}(0)\bigr)^{2}\rangle=R_{N}^{2}
f(t/R_{N}^{d_{w}})\,, \label{Diff-ScLaw}
\end{equation}
where $\mathbf{r}(t)$ is the position of the walker (here, the monkey) at time
$t$, $R_{N} \sim N^{1/d_{f}}$ is the gyration radius of the fractal with mass
(number of sites) $N$ and $f$ is a scaling function with the properties
\begin{equation}
f(x)\sim\left\{
\begin{array}
[c]{lll}
1 & \text{for} & x\rightarrow\infty\\
x^{2/d_{w}} & \text{for} & x\rightarrow0
\end{array}
\right.  \,. \label{ScFuProp}
\end{equation}
As a result of Einstein's relation $d_{w}=d_{f}+d_{R}$ for the fractal
dimension of the random walk, it follows that \cite{HaDjMaStWe84,HaBAv87}
\begin{equation}
d_{w}=d_{f}+d_{\min}\,. \label{EinstRel}
\end{equation}

As mentioned above, the {\em monkey on a tree} problem has been around under a
different name for some 30 years now. For background, we refer to the review
article on diffusion in disordered media by Havlin and Ben-Avraham
\cite{HaBAv87}. In recent years, significant advancements have been made in
numerical simulations on problems different from but closely related to the
{\em monkey on a tree} problem. A sophisticated Monte Carlo algorithm has been
used to simulate lattice animals and trees in 2 to 9 dimensions~\cite{Gra97}
and to measure their static scaling exponents with high
precision~\cite{HsNaGr05,HsGra11}. Furthermore, simulations have been performed
to determine with high precision the fractal and multifractal dimensions of
SAWs on percolations clusters in $2$ to $4$
dimensions~\cite{BlaJa2008,BlaJa2009}. Hence, we feel that state of the art
simulations of diffusion and transport and lattice animals in dimensions
suitable for reliable comparison to field theory have become within reach.
Thus, we think it is worthwhile to take a fresh look at the {\em monkey on a
tree} problem with field theoretic methods.

The static fractal dimension $d_f = 1/\nu_{\text{P}}$ of the randomly branched
polymers or trees are well known \cite{PaSo81,JaSt10}. Here, we apply
renormalized dynamical field theory to calculate $d_{\min}$ and the related
exponents in an $\varepsilon$-expansion to 2-loop order. Since $d_{\min}$ is
equal to the dynamical exponent $z$, the scaling exponent of the time a monkey
needs to reach a coconut on a tree, we can and will calculate $d_{\min}$ via
calculating $z$ of a stochastic process that generates RBPs \cite{JaSt10}.

\section{Model and field theoretic approach}
\label{sec:modelAndApproach}

This section serves 2 purposes. First, we review the field theoretic model for
RBPs that we have developed recently~\cite{JaSt10}. Some of the steps involved
in its derivation are far from trivial, and its symmetry contents is rich and
interesting. Hence, we think it is worthwhile to review the model in some
detail. This will also have the benefit of making the present paper more
self-contained. Second, we discuss in broad terms the diagrammatics resulting
from our dynamical model for the swollen phase and the collapse transition. We
will make the observation that the dynamical self-energy diagrams decompose
into a quasi-static part and a SAW-part that contains all the frequency
dependence. This observation is a key to the subsequent sections as it
simplifies the dynamical field theory for the transport and diffusion exponents
considerably.

\subsection{Creation of randomly branched polymers}

Our field theoretic model for RBPs is based on the idea of generating their
statistics through a mesoscopic stochastic growth process. It is well known
that the general epidemic process (GEP) \cite{Mol77} leads to random structures
with the properties of percolation clusters \cite{Gra83,Ja85,CaGra85,JaTa05}
which are, depending on the parameter-values of the GEP, below, at, or above
the percolation point. The primary density-fields describing this process are
the field of agents $n(\mathbf{r},t)$ and the field of the inactive debris
$m(\mathbf{r}
,t)=\lambda\int_{-\infty}^{t}dt^{\prime}\,n(\mathbf{r},t^{\prime})$ which
ultimately forms the polymer cluster. The following extension of the GEP is a
modification of a process that we have introduced for the description of
tricritical isotropic percolation \cite{JaMuSt04}. The non-Markovian Langevin
equation describing such this process (or rather its universality class) is
given by
\begin{equation}
\lambda^{-1}\partial_{t}n=\Big(\nabla^{2}+c\nabla m\cdot\nabla
\Big)n-\Big(r_{p}+g^{\prime}m+\frac{f^{\prime}}{2}m^{2}\Big)n+\zeta\,.
\label{Lang-Eq}
\end{equation}
Here, the parameter $r_{p}$ tunes the "distance" to the percolation threshold.
Below this threshold, in the absorbing phase, $r_{p}$ is positive which we
assume throughout this paper. In this case the typical final clusters of the
debris generated from a source $q\delta(\mathbf{r})\delta(t)$ of agents
consists of $N=\langle\int d^{d}r\,m(\mathbf{r},\infty)\rangle\approx q/r_{p}$
particles of the debris, and has a mean diameter (gyration radius)
$\sim1/\sqrt{r_{p}}$. However, here we are {\em not} interested in these
typical clusters. Rather, we are interested in the large non-typical clusters,
the rare events of the stochastic process, with $N\gg q/r_{p}$ (in this sense
$q$ is a small parameter). We know from percolation theory \cite{StAh94} that
these clusters belong to the universality class of lattice animals. Hence, they
are the same in a statistical sense as randomly branched polymers (RBPs) as far
as their universal properties go. The gradient-term proportional to $c$
describes the attractive influence of the debris on the agents if $c$ is
negative (as a negative contribution to $g^{\prime}$ does). In principle, other
gradient-terms like $m\nabla^{2}n$ and $n\nabla^{2}m$ could be added to the
Langevin equation. However as long as we have any one of these gradient terms
into our theory, an omission of the others has no effect on the final results,
and we choose to work with the term proportional to $c$ only for simplicity.
For usual percolation problems (ordinary or tricritical), all of these gradient
terms are irrelevant. As long as $g^{\prime}>0$, the third order term
$f^{\prime}m^{2}n$ is irrelevant near the transition point and the process
models ordinary percolation near $r_{p}=0$ \cite{Ja85} or non-typical very
large clusters, the swollen RBPs, for $r_{p}>0$. We permit both signs of
$g^{\prime}$ (negative values of $g^{\prime}$ correspond to an attraction of
the agents by the debris, see above). Hence, our model allows for a tricritical
instability (tricritical percolation near $r_{p}=0$ \cite{JaMuSt04} or the
collapse transition of the RBPs for $r_{p}>0$ \cite{JaSt10}). Consequently we
need the third order term with $f^{\prime}>0$ (representing self-avoidance) to
limit the density to finite values in these cases. Physically it originates
from the suppression of agents by the debris. The Gaussian noise-source
$\zeta(\mathbf{r},t)$ has correlations
\begin{align}
\overline{\zeta(\mathbf{r},t)\zeta(\mathbf{r}^{\prime},t^{\prime})}  &
=\Big(\lambda^{-1}gn(\mathbf{r},t)\delta(t-t^{\prime})-fn(\mathbf{r}
,t)n(\mathbf{r^{\prime}},t^{\prime})\Big)\nonumber\\
&  \qquad\qquad\qquad\times\delta(\mathbf{r}-\mathbf{r}^{\prime})\,.
\label{Abs-noise}
\end{align}
The first part of the noise correlation (\ref{Abs-noise}) takes into account
that the agents can decay spontaneously, and thus $g>0$. The term proportional
to $f$ simulates the anticorrelating or correlating (from attraction) behavior
of the noise in regions where debris has already been produced. If the coupling
constant $f$ becomes negative, attraction effects prevail. For ordinary
percolation this term is irrelevant. The form of the dependence of both parts
of the noise correlation on the field is mandated by the fact that the process
has to be strongly absorbing in order to model RBPs.

In the past, there have been misconceptions about the relation between
absorptivity and the form of the noise correlation, and we think that it is
worthwhile to address this relation here in a little more detail. As we just
mentioned, our RBP-generating process is like all percolation processes
strongly absorbing, i.e., its extinction probability (probability that the
process becomes extinct in a {\em finite} time-interval $(0,t)$) is larger than
zero, $P_{\text{ext}}(t)> 0$ for $t<\infty$. To guarantee this property, all
terms in the equation of motion (\ref{Lang-Eq}) must contain of course at least
one power of $n$. Moreover, and this is the important point here, the expansion
of the noise-correlation function~(\ref{Abs-noise}) in the density $n$ {\em
must} begin with a {\em linear} term \cite{Ja05}. With respect to enforcing
strong absorptivity, a quadratic part in $n$ of the noise correlation is
insufficient. The upshot is that phenomenological considerations are
sufficient to uniquely determine the form of the noise. There is no need to
resort to a microscopic formulation in terms of master
equations or the like to figure out the proper relevant
contributions to the noise correlations for the current process nor is it for
percolation processes in general.

\subsection{Field theoretic functionals}

To proceed towards a field theoretic model (for the general method of field
theory in statistical physics see, \textit{e.g.}, \cite{Am84,ZJ02}), the
Langevin equations (\ref{Lang-Eq}) and (\ref{Abs-noise}) are now transformed
into a stochastic response functional in the Ito-sense
\cite{Ja76,DeDo76,Ja92,JaTa05}
\begin{align}
\mathcal{J}_{\text{eGEP}}[\tilde{n},n]  &  =\int d^{d}x\Big\{\lambda\int
dt\,\tilde{n}\Big(\lambda^{-1}\partial_{t}-\nabla^{2}
\nonumber\\
& -c\nabla m\cdot \nabla +r_{p}+g^{\prime}m
+\frac{f^{\prime}}{2}m^{2}-\frac{g}{2}\tilde{n}\Big)n
\nonumber\\
&+\frac{f} {2}\Big(\lambda\int dt\,\tilde{n}n\Big)^{2}\Big\}\,. \label{StochFu}
\end{align}
where $\tilde{n}(\mathbf{r},t)$ denotes the (imaginary) response field
conjugated to $n(\mathbf{r},t)$. With this functional, we now have a vantage
point for the calculation of statistical quantities via path-integrals with the
exponential weight $\exp(-\mathcal{J}_{\text{eGEP}})$. When a source-term
$(\tilde{j},\tilde{n})$ is added, where
$\tilde{j}(\mathbf{r},t)=q\delta(\mathbf{r})\delta(t)$ and $(..,..)$ denotes an
integral of a product of two fields over space and time, this functional
describes, in particular, the statistics of clusters of debris generated by the
stochastic growth process (\ref{Lang-Eq}) from a source of $q$ agents at the
point $\mathbf{r}=0$ at time zero. Denoting by $\operatorname*{Tr}\bigl[\ldots
\bigr]$ the functional integration over the fields with boundary conditions
$n(\mathbf{r},-\infty)=\tilde{n}(\mathbf{r},\infty)=0$, we generally have for
the generating functional
\begin{equation}
\mathcal{Z}[j,\tilde{j}]=\operatorname*{Tr}\bigl[\exp\bigl(-\mathcal{J}
_{\text{eGEP}}[\tilde{n},n]+(\tilde{j},\tilde{n})+(j,n)\bigr]=1
\end{equation}
if the arbitrary sources $j$ {\em or} $\tilde{j}$ are zero. The first property
follows from causality whereas the second one originates from the absorptive
properties of the process. Note that the role of causality and adsorptivity can
be interchanged by the duality transformation
$m(\mathbf{r},t)\longleftrightarrow -\tilde{n}(\mathbf{r},-t)$
\cite{Ja85,Ja05,JaTa05}.

Averaging an observable $\mathcal{O}[n]$ over final clusters of debris (the
RBPs) of a given mass $N$ generated from the particular source
$\tilde{j}(\mathbf{r}
,t)=q\delta(\mathbf{r})\delta(t)$ leads to the quantity \cite{Ja85,JaTa05,Ja05}%
\begin{align}
&  \langle\mathcal{O}\rangle_{N}\mathcal{P}(N)=\left\langle \mathcal{O}%
[n]\delta(N-\mathcal{M})\exp\bigl((\tilde{j},\tilde{n})\bigr)\right\rangle
_{\text{eGEP}}\nonumber\\
&  =\operatorname*{Tr}\Big[\mathcal{O}[n]\delta(N-\mathcal{M})\exp
\bigl(-\mathcal{J}_{\text{eGEP}}+q\tilde{n}(\mathbf{0},0)\bigr)\Big]\nonumber\\
&  \simeq q\operatorname*{Tr}\Big[\mathcal{O}[n]\tilde{n}(\mathbf{0},0)\delta
(N-\mathcal{M})\exp\bigl(-\mathcal{J}_{\text{eGEP}}\bigr)\Big]\,,
\label{Erw-Wert}%
\end{align}
where
\begin{equation}
\mathcal{P}(N)=\langle\delta(N-\mathcal{M})\exp\bigl(q\tilde{n}%
(\mathbf{0},0)\bigr)\rangle_{\text{eGEP}} \label{defP}%
\end{equation}
is the probability distribution for finding a cluster (a RBP) of mass $N$.
\begin{equation}
\mathcal{M}=\int d^{d}rdt\,\lambda n(\mathbf{r},t)=\int d^{d}r\,m_{\infty
}(\mathbf{r})\, \label{Masse}%
\end{equation}
is the total mass of the debris. The field $m_{\infty}(\mathbf{r}%
)=m(\mathbf{r},t=\infty)$ describes the distribution of the debris after the
growth process terminated. Since the probability distribution should be
proportional to the number of different configurations, we expect by virtue of
universality arguments the following proportionality between the
probability distribution $\mathcal{P}(N)$ and the animal number $\mathcal{A}%
_{N}$ for asymptotically large $N$:
\begin{equation}
\mathcal{P}(N)\sim N\kappa_{0}^{-N}\mathcal{A}_{N}\sim N^{1-\theta}p_{0}%
^{N}\,, \label{A_zu_P}%
\end{equation}
where $\kappa_{0}$ and $p_{0}$ are non-universal in contrast to the universal
\textquotedblleft entropic\textquotedblright\ scaling exponent $\theta$. The
factor $N$ in Eq.~(\ref{A_zu_P}) arises because the generated clusters are
rooted at the source at the point $\mathbf{r}=0$, and each site of a given
animal may be the root of given cluster.

In actual calculations, the delta function appearing in averages like in
Eqs.~(\ref{Erw-Wert}) and (\ref{defP}) is hard to handle. This problem can be
simplified by using Laplace-transformed observables which are functions of a
variable conjugate to $N$, say $z$, and applying inverse Laplace transformation
(where all the singularities of the integrand lie to the left of the
integration path) in the end.  The switch to Laplace-transformed observables
can be done in a pragmatic way by augmenting the original
$\mathcal{J}_{\text{eGEP}}$ with a term $z\mathcal{M}$ and then working with
the new response functional
\begin{equation}
\mathcal{J}_{z}=\mathcal{J}_{\text{eGEP}}+z\mathcal{M}\,. \label{Def_Jz}%
\end{equation}
As an example, let us consider
\begin{align}
\langle\mathcal{O}\rangle_{N}\mathcal{P}(N)&=\int_{\sigma-i\infty}%
^{\sigma+i\infty}\frac{dz}{2\pi i}\,\mathrm{e}^{zN}
\nonumber\\
&\times\big\langle\mathcal{O}%
[n]\exp\bigl(-z\mathcal{M+}q\tilde{n}(0,0)\bigr)\big\rangle_{\text{eGEP}}\,.
\label{Inv-Lapl}%
\end{align}
Note that the relationship between $\mathcal{P}(N)$ and $\mathcal{A}_{N}$ given
in Eq.~(\ref{A_zu_P}) signals the existence of a singularity
$\sim(z-z_{c})^{\theta-2}$ of the integrand in Eq.~(\ref{Inv-Lapl}) at some
critical value $z_{c}$. Denoting averages with respect to the new functional by
$\langle\ldots \rangle_{z}$, and defining
\begin{equation}
q\Phi(z)=\ln\langle\exp(q\tilde{n})\rangle_{z}\approx q\langle\tilde{n}%
\rangle_{z} \label{Phi}%
\end{equation}
asymptotically, we get by using Jordans lemma that the asymptotic behavior,
\textit{e.g.,}\ of $\mathcal{P}(N)$ for large $N$ is given by
\begin{align}
\mathcal{P}(N)  &  =\int_{\sigma-i\infty}^{\sigma+i\infty}\frac{dz}{2\pi
i}\,\exp\bigl[zN+q\Phi(z)\bigr]\nonumber\\
&  \approx q\mathrm{e}^{z_{c}N+q\Phi(z_{c})}\int_{0}^{\infty}dx\,\frac
{\operatorname*{Disc}\Phi(z_{c}-x)}{2\pi i}\mathrm{e}^{-xN}\,,
\label{Probab_N}%
\end{align}
where the last row gives the asymptotics for large $N$. Here, $z_{c}$ is the
first singularity of $\Phi(z)\sim(z-z_{c})^{\theta-2}$, which is a branch point
on the negative real axis, and the contour of the path integral is deformed
into a path above and below the branch cut beginning at the singularity.
$\operatorname*{Disc}\Phi$ denotes the discontinuity of the
function $\Phi$ at the branch cut. The non-universal factor $q\mathrm{e}%
^{z_{c}N+q\Phi(z_{c})}$ depending exponentially on $N$ is common to all
averages defined by Eq.~(\ref{Erw-Wert}) and therefore cancels from all mean
values $\langle\mathcal{O}\rangle_{N}$.

Now, we return to our response functional $\mathcal{J}_{z}$ to refine it into a
form that suits us best for our actual field theoretic analysis. As discussed
above, the gradient term proportional to $c$ is redundant. To eliminate this
term, we apply to the field $\tilde{n}$ the shift and mixing transformation
\begin{equation}
\tilde{n}(\mathbf{r},t)\rightarrow\tilde{n}(\mathbf{r},t)+ a - ac
m_{\infty}(\mathbf{r})\,, \label{Shift}
\end{equation}
where $a$ is a free parameter at this stage. Ultimately, this parameter is
defined by $\langle\tilde{n}\rangle=0$ which means that the diagrammatic
perturbation expansion is free of tadpoles. Defining $\tau=r_{p}-ga$,
$\rho=(g^{\prime }+fa)a-ac\tau$, $h=z+r_{p}a-ga^{2}/2$, the stochastic
functional $\mathcal{J}_{z}$ (\ref{Def_Jz}) takes the form
\begin{align}
\mathcal{J}_{z}  &  =\int d^{d}x\Big\{\lambda\int dt\,\tilde{n}\Big(\lambda
^{-1}\partial_{t}+\tau-\nabla^{2}+g_{2}^{\prime}m-\frac{g_{2}}{2}\tilde
{n}\nonumber\\
&  +g_{1}m_{\infty}\Big)n+\Big(\frac{\rho}{2}m_{\infty}^{2}+\frac{g_{0}}%
{6}m_{\infty}^{3}+hm_{\infty}\Big)\Big\}\,. \label{Jz}%
\end{align}
Here, we could have set $\tau$ equal to zero by exploiting that $a$ is a free
parameter. Instead of doing so, we rather keep $\tau$ in our theory as a small
free redundant parameter. We will see later on that keeping $\tau$ comes in
handy for renormalization purposes. In Eq.~(\ref{Jz}), we have eliminated
couplings that are of more than third order in the fields because they are
irrelevant. We do not write down in detail the relatively uninteresting
relations between the new third-order coupling constants and the old ones. Note
that $\mathcal{J}_{z}$ contains two similar couplings: $g_{2}^{\prime
}\tilde{n}nm$ and $g_{1}\tilde{n}nm_{\infty}$. Whereas the first coupling
respects causal ordering, which means that $\tilde{n}$ is separated by an
infinitesimal positive time-element from the $nm$-part resulting from the
Ito-calculus \cite{Ja92}, the second one respects causality only between
$\tilde{n}$ and $n$. In contrast to the $m$-part, the $m_{\infty}$-part
contains all the $n$ with times that lie in the past and in the future of
$\tilde{n}$. This property is the heritage of the time-delocalized noise term,
and of the introduction of the $\delta$-function in Eq.~(\ref{Erw-Wert}) as a
final ($t=\infty$) condition that destroyed the causality of $\mathcal{J}_{z}$.
Even if we had disregarded the noise term proportional to $f$ in
Eq.~(\ref{Abs-noise}) initially, the $\tilde{n}nm_{\infty}$-coupling would be
generated by coarse graining, and hence it must be ultimately incorporated into
the theory to yield renormalizability.

The relevance or irrelevance of the different terms in $\mathcal{J}_{z}$
follows from their dimensions with respect to an inverse length scale $\mu$
such that time scales as $\mu^{-2}$. Fundamentally, one has to decide which
parameters are the critical control-parameters going to zero in mean-field
theory. At the collapse transition these are $\tau\sim\rho\sim\mu^{2}$, and
$h\sim \mu^{(d+2)/2}$ \cite{JaSt10}. The dimensions of the fields are then
given by $\tilde{n}\sim m\sim\mu^{(d-2)/2}$, and $n\sim\mu^{(d+2)/2}$. It
follows that all the coupling constants $g_{0}$, $g_{1}$, $g_{2}$, and
$g_{2}^{\prime}$ have the same dimension $\mu^{(6-d)/2}$. Note that $\tilde{n}$
is tied always to at least one factor of $n$ as a result of absorptivity of the
process. Hence, all the terms in $\mathcal{J}_{z}$ are relevant for $d\leq6$
spatial dimensions, and the upper critical dimension of the collapse transition
is $d_{c}=6$. The situation is different if $\rho$ is a finite positive
quantity, that is in the swollen phase. Then $\rho$ can be absorbed into the
fields by a rescaling transformation which amounts to formally setting
$\rho=2$. The field dimensions then become $m\sim\mu^{d/2}$, $n\sim
\mu^{(d+4)/2}$, and $\tilde{n}\sim\mu^{(d-4)/2}$. It follows that $h\sim
\mu^{d/2}$, $g_{0}\sim\mu^{-d/2}$, $g_{1}\sim\mu^{(4-d)/2}$, $g_{2}^{\prime
}\sim\mu^{(2-d)/2}$, and $g_{2}\sim\mu^{(8-d)/2}$. Hence, in the swollen phase
only $g_{2}=g$ is relevant, now below $8$ spatial dimensions. The other
couplings can be safely removed. Then, $\mathcal{J}_z$ reduces to the dynamical
response functional of the usual simple GEP.

Recently, we have shown that $g_{2}^{\prime}$ becomes weakly irrelevant (with a
correction exponent of order $O(6-d)$) at the RG fixed point corresponding to
the collapse transition of RBPs~\cite{JaSt10}. Hence, we can neglect this
coupling not only for the swollen phase but also for the collapse transition,
even though dimensional analysis suggests its relevance for the latter in the
vicinity of the Gaussian fixed point. In Ref.~\cite{JaSt10} we argued that hat
the vanishing of $g_{2}^{\prime}$ indicates that cycles or loops of RBPs are
irrelevant. In other words: large RBPs in the swollen phase and at the collapse
transition are dominated by tree-configurations.

The last step in setting up the response functional for the dynamical
description of swollen or collapsing RBPs is a duality transformation that
interchanges absorptivity in favor of causality:
\begin{align}
\tilde{n}(\mathbf{r},t)  &  =s(\mathbf{r},-t)\,,\qquad n(\mathbf{r}%
,t)=\tilde{s}(\mathbf{r},-t)\,,\nonumber\\
m(\mathbf{r},t)  &  =\lambda\int_{-t}^{\infty}dt^{\prime}\,\tilde
{s}(\mathbf{r},t^{\prime})\,,\nonumber\\
m_{\infty}(\mathbf{r}) & = m(\mathbf{r},\infty)=:\tilde
{\varphi}(\mathbf{r}). \label{Abs_Caus}%
\end{align}
This step leads to the response functional
\begin{align}
\mathcal{J}  &  =\int d^{d}x\Big\{\lambda\int dt\,\tilde{s}\Big(\lambda
^{-1}\partial_{t}+\tau-\nabla^{2}+g_{1}\tilde{\varphi}-\frac{g_{2}}%
{2}s\Big)s\nonumber\\
&  \qquad\qquad\qquad\qquad+\Big(\frac{\rho}{2}\tilde{\varphi}%
^{2}+\frac{g_{0}}{6}\tilde{\varphi}^{3}+h\tilde{\varphi}\Big)\Big\}\,,
\label{J-An}%
\end{align}
which will serve us in the following as the vantage point of our dynamical
field theory.

In this paper, we are mainly interested in the dynamical aspects of our theory.
On occasion, however, we will also discuss some of its static aspects. On one
hand, this will make our presentation more self contained because we will need
some of the previously derived static results as input as we move along. On the
other hand, we feel that certain elements of the theory are more easily
discussed statically rather than dynamically. For the static aspects, we do not
need the full response functional. Rather it is sufficient to consider the
quasi-static Hamiltonian
\begin{align}
\mathcal{H}  &  =\int d^{d}x\Big\{\tilde{\varphi}\bigl(\tau-\nabla
^{2}\bigr)\varphi+\frac{\rho}{2}\tilde{\varphi}^{2}+h\tilde{\varphi}\nonumber\\
&\qquad\qquad+\frac{g_{0}}{6}\tilde{\varphi}^{3}+g_{1}\tilde{\varphi}^{2}\varphi
-\frac{g_{2}}{2}\tilde{\varphi}\varphi^{2}\nonumber\\
&\qquad\qquad+\bar{\psi}\bigl(\tau-\nabla^{2}
+g_{1}\tilde{\varphi}-g_{2}\varphi\bigr)\psi\Big\}\,, \label{H-An}
\end{align}
that follows from $\mathcal{J}$ by setting
$s(\mathbf{r},t)\rightarrow\varphi(\mathbf{r})$. Here, we have added a pair of
fermionic ghost fields $(\bar{\psi},\psi)$ that automatically guarantees the
original causality rule in Feynman diagrams through its couplings to the fields
$\tilde{\varphi}$ and $\varphi$ \cite{JaSt10}. We note that the quasi-static
Hamiltonian has BRS supersymmetry which indicates that only tree-like polymer
configurations are relevant as noted above.

\subsection{Diagrammatics: SAWs on tree diagrams}
\label{subsec:diagrammatics}

\begin{figure}[ptb]
\centering{\includegraphics[width=6cm]{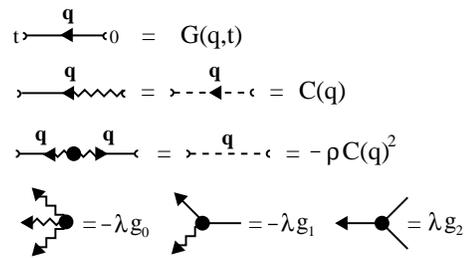}}\caption{The elements of the
dynamic Feynman diagrams}%
\label{fig:elements}%
\end{figure}

Figure~\ref{fig:elements} shows the diagrammatic elements of our theory as
resulting from $\mathcal{J}$ and $\mathcal{H}$. There is the dynamic propagator
which reads
\begin{equation}
G(\mathbf{q},t)=\theta(t)\,\exp\bigl[-(\tau+\mathbf{q}^{2})t\bigr]\,, \label{Prop-1}%
\end{equation}
in momentum-time representation or, after Fourier transformation,
\begin{equation}
\tilde{G}(\mathbf{q},\omega)=\frac{1}{i\lambda\omega+\tau+\mathbf{q}^{2}}\,,
\label{Prop-1-fr}%
\end{equation}
in momentum-frequency representation. Its static counterpart is given by
\begin{equation}
C(\mathbf{q})=\tilde{G}(\mathbf{q},0)=\frac{1}{\tau+\mathbf{q}^{2}}\,. \label{Prop-2}%
\end{equation}
The dynamical functional also features the static correlator
\begin{equation}
-\rho C(\mathbf{q})^{2}=\frac{-\rho}{\bigl(\tau+\mathbf{q}^{2}\bigr)^{2}}\,.
\label{Correl}%
\end{equation}
Note the negative sign of the correlator that has its origin in the fact that
the fluctuations of the fields $s$ and $\varphi$ are purely imaginary (a
heritage of the imaginary response field $\tilde{n}$). In addition to these
elements represented by lines, there are the 3 vertices shown at the bottom of
Fig.~\ref{fig:elements}.

\begin{figure}[ptb]
\centering{\includegraphics[width=4cm]{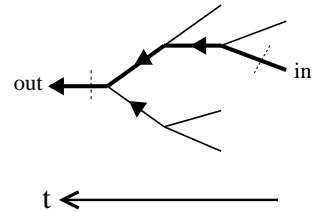}}\caption{A tree-like subdiagram.}%
\label{fig:tree}%
\end{figure}

\begin{figure}[ptb]
\centering{\includegraphics[width=4cm]{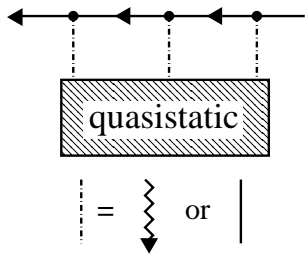}}\caption{Decomposition of a
self-energy diagram into a dynamic connecting SAW and a quasi-static part.}%
\label{fig:quasimean}%
\end{figure}

Recall that we are primarily interested in scaling exponents for transport and
diffusion on RBPs. These, we will extract from the response function
\begin{equation}
G_{1,1}(\mathbf{r},t)=\langle
n(\mathbf{r},t)\tilde{n}(\mathbf{0},0)\rangle=\langle s(-\mathbf{r}
,t)\tilde{s}(\mathbf{0},0)\rangle
\end{equation}
and its renormalizations. Its (quasi-)static renormalizations are well known
from earlier work \cite{PaSo81,AKM81,JaSt10}. Its dynamical renormalization is
known only to 1-loop order for the swollen phase as shown by one of us
\cite{Ja85} and entirely unknown for the collapse transition. Hence, it is our
main task here to calculate the renormalization factor $Z_\lambda$ pertaining
to the dynamical coefficient $\lambda$. To this end, we consider the
one-particle irreducible amputated self-energy diagrams with an outgoing
amputated $\tilde{s}$-leg and an ingoing amputated $s$-leg. Let us forget for
the moment all the static $C$-lines of a diagram that are introduced by the
couplings to the static field $\tilde{\varphi}$. Then any of these diagrams is
reduced to a pure time-ordered tree-diagram that has its origin in the outgoing
amputated $\tilde{s}$-leg, see Fig.~\ref{fig:tree}.
The tree consists of $G$-lines and $g_{2}$-vertices. The ingoing amputated
$s$-leg is one of the pending $s$-legs. The SAW along $G$-lines which connects
the two amputated legs is unique. Note, that the tree-structure with this
unique SAW is a consequence of the limit $g_{2}^{\prime}\rightarrow0$, which we
may hence call the tree-limit. Now we reintroduce the $C$-lines by inserting
$g_{0}$-, $g_{1}$- and $\rho$-vertices and saturating the pending non-amputated
$s$-legs with the static $\tilde{\varphi}$-legs. Then, integrating over
internal times from right (earlier times) to left (later times), it is easy to
see that the internal $\tilde{s}$-legs are converted by the integrations into
$\tilde{\varphi}$-legs with exception of the $\tilde{s}$ that are part of the
connecting SAW between the amputated external two legs. The conversion of
$\tilde{s}$-legs into $\tilde{\varphi}$-legs turns the corresponding $G$-lines
into $C$-lines. In other words, the parts of the diagram that do not belong to
the connecting SAW become purely quasi-static. One can think of the
time-integrations as having the net effect of decomposing any self-energy
diagram into a dynamic part, the connecting SAW, and a quasi-static residual
part. This decomposition is visualized in Fig.~\ref{fig:quasimean}. After the
decomposition, the frequency dependence of any self-energy diagrams solely
resides in its connecting SAW. To calculate the dynamical renormalization
factor $Z_\lambda$, we have to calculate the parts of the self-energy diagrams
that are proportional to $i\omega$ which we can do, following standard field
theoretic procedures, by making $i\omega s \tilde{s}$-insertions. After what we
just have learned about the decomposition effect of the time integrations, it
is clear need to put these insertions only into the $G$-lines forming the
connecting SAW.

All these considerations are easily generalized from self-energy diagrams to
general  one-particle irreducible diagram with $\tilde{N}$ external
$\tilde{s}$-legs and ${N}$ external ${s}$-legs. Let us consider such a diagram
beginning, as above, by temporarily omitting all static $C$-lines. This leads
to a decomposition of the diagram into a (spanning) forest of $\tilde{N}$
disconnected time-directed trees rooted in the outgoing amputated
$\tilde{s}$-legs each featuring a subset of the $N$ incoming amputated
$s$-legs. Then,  Fourier-transformation from time-arguments to frequencies
shows that the frequency flowing out through a $\tilde{s}$-leg is the sum of
the frequencies flowing into the diagram through the corresponding set of
$s$-legs. Thus, there are $\tilde{N}$ frequency-conservation laws. The argument
concludes by tying together the disconnected trees by readmitting the static
$C$-lines.

\section{The swollen RBP}
\label{sec:swollenRBP}

In this section, we use our model and our insight into the structure of the
diagrammatic expansion developed in Sec.~\ref {subsec:diagrammatics} to
calculate the dynamical exponent $z$ and the related transport exponents for
the swollen phase. Before we embark on this quest, we will briefly review some
of the quasi-static results that we can utilize as input for our dynamical
calculation. Furthermore, we have performed rational approximations to improve
the known results for the static exponents that we would like to present here
along with comparisons to the available numerical results for these exponents.

\subsection{Functionals, renormalizations, and static results}

As discussed above, $g_0$, $g_1$, and  $g_2^\prime$ are strongly irrelevant for
the swollen phase, and hence we now set $g_{0}=g_{1}=g_2^\prime=0$. Moreover,
$\rho$ is a finite and positive quantity that can be reset through a simple
re-scaling transformation. Hence, we have the freedom to choose $\rho=2$ for
simplicity and we do so.
With these settings, the response functional reduces to%
\begin{align}
\mathcal{J}_{sw}  &  =\int d^{d}x\Big\{\lambda\int dt\,\tilde{s}%
\bigl(\lambda^{-1}\partial_{t}+\tau-\nabla^{2}-\frac{g}{2}s\bigr)s\nonumber\\
&  \qquad\qquad\qquad\qquad\qquad+\bigl(\tilde{\varphi}^{2}+h\tilde{\varphi
}\bigr)\Big\}\,, \label{J-swAn}%
\end{align}
where $g = g_2$.This functional describes besides the dynamical creation of the
swollen RBPs the dynamics at the Yang-Lee singularity \cite{BrJa81}, however
with quenched static noise \cite{Ja85}. With the settings for the swollen
phase, the quasi-static Hamiltonian including ghost-fields becomes
\begin{align}
\mathcal{H}_{sw}  &  =\int d^{d}x\Big\{\tilde{\varphi}\bigl(\tau-\nabla
^{2}-\frac{g}{2}\varphi\bigr)\varphi+\tilde{\varphi}^{2}+h\tilde{\varphi
}\nonumber\\
& \qquad\qquad\qquad
+\bar{\psi}\bigl(\tau-\nabla^{2}-g\varphi\bigr)\psi\Big\}\nonumber\\
=& \int d^{d}x d\theta d\bar{\theta}\,\Big\{\frac{1}
{2}\Phi\Big(\tau-\square\Big)\Phi-\frac{g}{6}\Phi^{3}+h\Phi\Big\}\,,
\label{H-swAn}
\end{align}
where we have introduced Grassmannian anticommuting super-coordinates $\theta$,
$\bar{\theta}$ with integration rules $\int d\theta\,1=0$, $\int d\bar{\theta
}\,1=0$, $\int d\theta\,\theta=1$, $\int d\bar{\theta}\,\bar{\theta}=1$, and
defined a super-field $\Phi(\mathbf{r},\bar{\theta},\theta)=\varphi
(\mathbf{r})+i\bar{\theta}\psi(\mathbf{r})+i\bar{\psi}(\mathbf{r})\theta
+\bar{\theta}\theta\tilde{\varphi}(\mathbf{r})$, as well as the super-Laplace
operator $\square=\nabla^{2}+2\partial_{\bar{\theta}}\partial_{\theta}$. The
Hamiltonian $\mathcal{H}_{sw}$ shows full supersymmetry, i.e., besides the
symmetry against super-translations (BRS-symmetry) it also has super-rotation
symmetry. Parisi and Sourlas \cite{PaSo81} showed some 30 years ago that the
full supersymmetry leads to dimensional reduction because it makes the
Hamiltonian equivalent to the ordinary Yang-Lee-Hamiltonian in two lesser
dimensions (see also the rigorous work on the dimensional reduction of Brydges
and Imbrie \cite{BrIm03,Ca2003}). Exploiting this relation, all static
renormalizations are known up to third order \cite{AKM81}.

We note that in contrast to the quasi-static Hamiltonian which, as we just have
seen, can be written in the form of the supersymmetric Yang-Lee-Hamiltonian by
the introduction of a superfield, the dynamic response functional is not
supersymmetric. Hence, we unfortunately cannot exploit dimensional reduction in
our dynamic calculation.

In the following, we use dimensional regularization. For the swollen phase, we
employ the renormalization scheme
\begin{subequations}
\label{Ren-Sch}%
\begin{align}
s  &  \rightarrow\mathring{s}=Z^{1/2}s\,,\quad\tilde{s}\rightarrow
\mathring{\tilde{s}}=Z_{\lambda}Z^{1/2}\tilde{s}\,,
\\
\lambda &  \rightarrow\mathring{\lambda}=Z_{\lambda}^{-1}\lambda\,,\quad
\tau\rightarrow\mathring{\tau}=Z^{-1}Z_{\tau}\tau\,,
\\
g  &  \rightarrow\mathring{g}=Z^{-3/2}Z_{g}g\,,\quad G_{\varepsilon}g^{2}=
u\mu^{\varepsilon}\,,
\\
h  &  \rightarrow\mathring{h}=Z^{-1/2}\Big(h-\frac{g}{2}A \tau^{2}\Big)\, ,
\\
(\tilde{\varphi},\varphi,\bar{\psi},\psi)  &  \rightarrow(\mathring
{\tilde{\varphi}},\mathring{\varphi},\mathring{\bar{\psi}},\mathring{\psi
})=Z^{1/2}(\tilde{\varphi},\varphi,\bar{\psi},\psi)\,.
\end{align}
\end{subequations}
Not all the renormalization $Z$-factors in the renormalization scheme are
independent. The form-invariance  of $\mathcal{J}_{sw}$ and $\mathcal{H}_{sw}$
under the shift $s \to s + \alpha$ leads to the Ward identities
$Z_{g}=Z_{\tau}$ and $Z_{\tau}=1+g^{2}A$ \cite{BrJa81}. As mentioned above, the
$Z$-factors other than $Z_\lambda$ are known to 3-loop order. In our dynamical
calculation 2-loop calculation, we will exploit the known result for the field
renormalization $Z$ to 2-loop oder as an input
\begin{align}
Z  &  =1+\frac{u}{3\varepsilon}+\Big(5-\frac{13}{12}\varepsilon\Big)\Big(\frac
{u}{3\varepsilon}\Big)^{2}+O(u^{3})\,,
 \label{Z-sw}%
\end{align}
where $\varepsilon=8-d$.

For completeness, let us mention here that the known $3$-loop result for the
entropic scaling exponent $\theta$ as featured in Eq.~(\ref{A_zu_P}) reads
\cite{AKM81}
\begin{equation}
\theta=\frac{5}{2}-\frac{\varepsilon}{12}-\frac{79}{3888}\varepsilon
^{2}+\Big(\frac{\zeta(3)}{81}-\frac{10445}{1259712}\Big)\varepsilon
^{3}+O(\varepsilon^{4})\,. \label{theta-eps}%
\end{equation}
The scaling exponent $\nu_{\text{P}}$ of the gyration radius of the branched
polymer, $R_{N}\sim
N^{\nu_{\text{P}}}$, is related to $\theta$ by%
\begin{equation}
\theta=(d-2)\nu_{\text{P}}+1\,, \label{theta-nu}%
\end{equation}
The resulting $\varepsilon$-expansion of $\nu_{\text{P}}$ reads
\begin{equation}
\nu_{\text{P}}=\frac{1}{4}+\frac{\varepsilon}{36}+\frac{29}{23328}%
\varepsilon^{2}+\Big(\frac{\zeta(3)}{486}-\frac{8879}{7558272}\Big)\varepsilon
^{3}+O(\varepsilon^{4})\,. \label{nu-eps}%
\end{equation}
Now, we improve these results by making rational approximations that
incorporate the exact results that are known for $\theta$ in dimensions $d$
from $1$ to $4$, namely $\theta(d=1)=0$, $\theta(d=2)=1$, $\theta(d=3)=3/2$,
$\theta (d=4)=11/6$.  We get
\begin{equation}
\theta\approx\frac{5}{2}-\frac{\varepsilon}{12}\Big(\frac{1+1.37622\varepsilon
-0.0130833\varepsilon^{2}-0.0171653\varepsilon^{3}}{1+1.13239\varepsilon
-0.210607\varepsilon^{2}+0.00685362\varepsilon^{3}}\Big)\,. \label{theta-rati}%
\end{equation}
For $\nu_{\text{P}}$, we include the exact values $\nu_{\text{P}}(d=1)=1$,
$\nu_{\text{P}}(d=3)=1/2$,
$\nu_{\text{P}}(d=4)=5/12$ and get%
\begin{equation}
\nu_{\text{P}}\approx\frac{1}{4}+\frac{\varepsilon}{36}\Big(\frac
{1+1.17714\varepsilon-0.113178\varepsilon^{2}}{1+1.13239\varepsilon
-0.210607\varepsilon^{2}+0.00685362\varepsilon^{3}}\Big)\,. \label{nu-rati}%
\end{equation}
With regard to the relation (\ref{theta-nu}), we note that our independent
rational approximations satisfy these relations to order $O(10^{-6})$. In
Table~\ref{tab:An-kritExp}, we compare the numerical values resulting from
these approximations for various dimensions to recent simulation results by
Hsu, Nadler, and Grassberger~\cite{HsNaGr05,HsGra11}. Over all, the two agree
remarkably well over a wide range of dimensions.
\begin{table*}[ptb]%
\caption{Rational approximation estimates of the exponents $\theta$ und
$\nu_{\text{P}}$ compared with simulation results by Hsu, Nadler, and
Grassberger \cite{HsNaGr05}.}
\begin{tabular}
[c]{c|cccccccc}\hline\hline $d$ & \quad$1$ & \quad$2$ & \quad$3$ & \quad$4$ &
\quad$5$ & \quad$6$ & \quad$7$ & \quad$8$\\\hline $\theta$ & \quad$0$ &
\quad$1$ & \quad$3/2$ & \quad$11/6$ & \quad$2.0769$ &
\quad$2.2603$ & \quad$2.3986$ & \quad$5/2$\\
\quad Hsu \emph{et al.} &  &  &  & $\quad1.835(6)$ & $\quad2.080(7)$ &
$\quad2.261(12)$ & $\quad2.40(2)$ & \\
$\nu_{\text{P}}$ & \quad$1$ & \quad$0.64267$ & \quad$1/2$ & \quad$5/12$ &
\quad$0.35896$ & \quad$0.31507$ & \quad$0.27973$ & \quad$1/4$\\
\quad Hsu \emph{et al.} &  & $\quad0.6412(5)$ &  & $\quad0.4163(30)$ &
$\quad0.359(4)$ & $\quad0.315(4)$ & $\quad0.282(5)$ & \\\hline\hline
\end{tabular}
\label{tab:An-kritExp}%
\end{table*}
\begin{figure}[ptb]
\centering{\includegraphics[width=6cm]{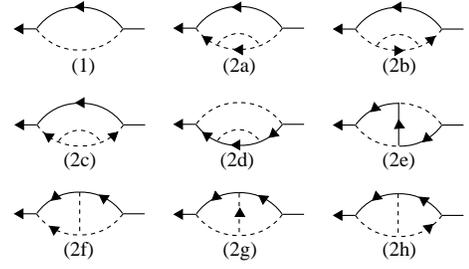}}\caption{Selfenergy diagrams
for swollen RBPs to 2-loop order.}%
\label{fig:sw-diagrams}%
\end{figure}

\subsection{Calculation of the dynamical scaling exponent}

Now we return to our main goal, the 2-loop calculation of the dynamical
exponent $z$ via the calculation of $Z_{\lambda}$. As a warmup, let us first do
a quick 1-loop calculation that reproduces the known 1-loop result. At this
order, there is only one self-energy diagram, namely diagram (1) of
Fig.~(\ref{fig:sw-diagrams}). Using the graphical elements pictured in
Fig.~(\ref{fig:elements}), this diagram translates into the formula
\begin{align}
(1)  &  =\lambda g^{2}\int_{\mathbf{p}}\frac{(-2)}{\bigl(\tau+\mathbf{p}%
^{2}\bigr)^{2}\bigl(i\omega/\lambda+\tau+(\mathbf{p}-\mathbf{q})^{2}%
\bigr)}\,,
\label{S(1-1)}%
\end{align}
where $\int_{\mathbf{p}}\ldots$ is an abbreviation for the $d$-dimensional
momentum integral $(2\pi)^{-d}\int d^{d}p\ldots$. The calculation of this and
in particular that of the higher order diagrams can be simplified be resorting
to massless propagators. Practically, this is done by expanding the integrands
in powers of $i \omega/\lambda$ and $\tau$. For the 1-loop diagram (1) of
Fig.~(\ref{fig:sw-diagrams}) this gives
\begin{align}
(1)  &   =-2\lambda g^{2}\int_{\mathbf{p}}\Bigg\{\frac{1}{(\mathbf{p}-\mathbf{q}%
)^{2}\bigl(\mathbf{p}^{2}\bigr)^{2}}-\frac{i\omega/\lambda+\tau}%
{\bigl((\mathbf{p}-\mathbf{q})^{2}\bigr)^{2}\bigl(\mathbf{p}^{2}\bigr)^{2}%
}
\nonumber\\
&-\frac{2\tau}{(\mathbf{p}-\mathbf{q})^{2}\bigl(\mathbf{p}^{2}\bigr)^{3}%
}+\ldots\Bigg\}
\,. \label{S(1-2)}%
\end{align}
Note that the massless integrals produced by the expansion are not infrared
divergent as long as $d>6$. For the calculation of these integrals, we refer to
Appendix~\ref{app:massless_calculation}. Their $\varepsilon$-expansions about
$d=8$ lead to the
renormalized vertex function to $1$-loop order%
\begin{align}
\Gamma_{1,1}^{(1L)}(q,\omega)  &  =\bigl(ZZ_{\lambda}i\omega+Z_{\tau}%
\lambda\tau+Z\lambda q^{2}\bigr)\nonumber\\
&  -\frac{u}{3\varepsilon}\bigl(2i\omega+6\lambda\tau+\lambda q^{2}%
\bigr)+\ldots\,. \label{Gamma-1L}%
\end{align}
Hence, we obtain to this order%
\begin{equation}
Z=Z_{\lambda}=1+\frac{u}{3\varepsilon}\,,\qquad Z_{\tau}=1+\frac
{2u}{\varepsilon}\,, \label{Z-1L}%
\end{equation}
which leads to the 1-loop result  $z=2+\eta+O(\varepsilon^{2})$~ \cite{Ja85},
where $\eta=-\varepsilon/9+O(\varepsilon^{2})$ is the anomalous dimension of
the field $s$ \cite{BrJa81}.

Now turn to the 2-loop part of the calculation for which we consider the
$2$-loop diagrams of Fig.~(\ref{fig:elements}). The parts of these diagrams
linear in $i\omega$ are
\begin{subequations}
\label{Diag-2loop}%
\begin{align}
&\Big((2a)+(2b)+(2c)\Big)_{i\omega} = \Big(2F(5,2,1)+\frac{1}{2}%
F(4,2,2)\Big)
\nonumber\\
&\qquad\qquad\qquad\qquad\qquad\qquad\quad\times \bigl(-4i\omega g^{4}\bigr)\,,
\\
&(2d)_{i\omega}  =\Big(2F(5,2,1)+F(4,2,2)\Big)\bigl(-4i\omega g^{4}\bigr)\,,
\\
&(2e)_{i\omega} = \Big(2F(4,3,1)+F(3,3,2)\Big)\bigl(-4i\omega g^{4}\bigr)\,,
\\
&\Big((2f)+(2g)+(2h)\Big)_{i\omega} = \Big(2F(4,3,1)+2F(4,2,2)
\nonumber\\
&\qquad\qquad\qquad\qquad\quad +2F(3,3,2)\Big)\bigl(-4i\omega g^{4}\bigr)\,,
\end{align}
\end{subequations}
where the $F(\ldots)$ are the frames of the diagrams, i.e., their parts
consisting only of the momentum integrations without any coupling constants or
symmetry factors. For a more precise definition of the $F(\ldots)$, we refer to
Appendix~\ref{app:massless_calculation}. Next, we calculate the counter-terms
of these frames. For the simplicity of the argument, let us just say here that
we apply some calculation procedure $\mathfrak{C}$ to the frames that produces
as its result the counter-terms of the frames:
\begin{align}
\mathfrak{C} \, F(\ldots)=C(\ldots) \label{procedure}
\end{align}
A precise definition of $\mathfrak{C}$ along with some details of the
calculations is given in Appendix~\ref{app:massless_calculation}. Application
of $\mathfrak{C}$ results in
\begin{align}
&\mathfrak{C}\Big((2a)+\cdots+(2h)\Big)_{i\omega}    =\Big(4C(5,2,1)+\frac
{7}{2}C(4,2,2)
\nonumber\\
&\qquad\qquad\qquad+4C(4,3,1)+3C(3,3,2)\Big) \bigl(-4i\omega g^{4}\bigr)\nonumber\\
&\qquad\qquad\qquad\qquad
=\frac{11}{9\varepsilon^{2}}\Big(1-\frac{43}{132}\varepsilon\Big)i\omega
u^{2}\,. \label{counter-2loop}%
\end{align}
These counter-terms yield the $2$-loop contribution to the
renormalization-factor product
\begin{equation}
ZZ_{\lambda}=1+\frac{2u}{3\varepsilon}+\frac{11}{9\varepsilon^{2}}%
\Big(1-\frac{43}{132}\varepsilon\Big)u^{2}+O(u^{3})\,. \label{ZZ_lambd}%
\end{equation}
Using Eq.~(\ref{Z-sw}), we finally obtain the wanted dynamic renormalization
factor
\begin{equation}
Z_{\lambda}=1+\frac{u}{3\varepsilon}+\frac{5}{9\varepsilon^{2}}\Big(1-\frac
{\varepsilon}{2}\Big)u^{2}+O(u^{3})\,. \label{Z_lambd}%
\end{equation}

\begin{table*}[ptb]%
\caption{Rational-approximation estimates of the fractal dimensions of swollen
RBPs compared with numerical results of Havlin, Djordjevic, Majid, Stanley, and
Weiss \cite{HaDjMaStWe84}.}
\begin{tabular}
[c]{c|cccccccc}\hline\hline $d$ & \quad$1$ & \quad$2$ & \quad$3$ & \quad$4$ &
\quad$5$ & \quad$6$ & \quad$7$ & \quad$8$\\\hline $d_{w}$ & \quad$2$ &
\quad$2.7138$ & \quad$3.3127$ & \quad$3.8638$ &
\quad$4.3960$ & \quad$4.9238$ & \quad$5.4560$ & \quad$6$\\
\quad Havlin \emph{et al.} &  & $\quad2.78(8)$ & \quad$\quad3.37(10)$ &
$\quad\quad3.89(12)$ & $\quad$ & $\quad$ & $\quad$ & \\
$d_{\min}$ & \quad$1$ & \quad$1.1578$ & \quad$1.3127$ & \quad$1.4638$ &
\quad$1.6101$ & \quad$1.7499$ & \quad$1.8811$ & \quad$2$\\
\quad Havlin \emph{et al.} &  & $\quad1.17(8)$ & $\quad\quad1.36(10)$ &
$\quad\quad1.49(12)$ & $\quad$ & $\quad$ &  & \\
$d_{f}$ & \quad$1$ & \quad$1.5560$ & \quad$2$ & \quad$12/5$ & \quad$2.7859$ &
\quad$3.1739$ & \quad$3.5749$ & \quad$4$\\\hline\hline
\end{tabular}
\label{tab:Fract-kritExp}%
\end{table*}

Next, we discuss the renormalization group equation for the vertex functions
and its solution. As the result of the dynamic tree structure of the Feynman
diagrams, the vertex functions
$\Gamma_{\tilde{N},N}(\{\omega/\lambda\},\{\mathbf{q}\},\tau,u,\mu)$ depend
only on the incoming frequencies of the $N$ $s$-legs. Setting these frequencies
to zero, we get in the case $\tilde{N}=1$ the static vertex functions
\begin{equation}
\Gamma_{1,N}^{(\text{stat})}(\{\mathbf{q}\})=\Gamma_{1,N}(\{\omega/\lambda=0\},
\{\mathbf{q}\}) \label{statGamma}
\end{equation}
that are related by dimensional reduction to the vertex functions
$\Gamma_{N+1}^{(YL)}$ of the Yang-Lee theory in two lesser dimensions. The RGE
of the dynamical $\Gamma_{\tilde{N},N}$ reads \cite{BrJa81}
\begin{equation}
\mathcal{D}_{\mu}\Gamma_{\tilde{N},N} =
\gamma\frac{(\tilde{N}+N)}{2}\Gamma_{\tilde{N},N} -
\gamma_{\tau}\frac{\tau^{2}}{2g}\,\delta_{\tilde{N},1}\delta_{N,0}\,,
\label{RGG-sw}
\end{equation}
where
\begin{equation}
\mathcal{D}_{\mu} = \mu\partial_{\mu}+
\zeta\lambda\partial_{\lambda}+\beta\partial _{u}+\kappa\tau\partial_{\tau}
\label{RG-Op}
\end{equation}
is the RG differential operator, and
\begin{align}
\gamma &  =\left.  \mu\partial_{\mu}\ln Z\right\vert_0 \,,\quad \gamma_{\tau
}=\left.  \mu\partial_{\mu}\ln Z_{\tau}\right\vert_0 \,,\quad\zeta=\left.
\mu\partial_{\mu}\ln Z_{\lambda}\right\vert_0 \,,\nonumber\\
\kappa & =\gamma-\gamma_{\tau}\,,\qquad\quad
\beta=(-\varepsilon+3\gamma-2\gamma_{\tau})u\,. \label{RGGfu-sw}
\end{align}
For the Wilson functions featured here, we have the expansions
\begin{align}
\gamma &  =-\frac{u}{3}+\frac{13}{54}u^{2}+O(u^{3})\,,\quad\gamma_{\tau
}=-2u+\frac{23}{6}u^{2}+O(u^{3})\,,\nonumber\\
\zeta &  =-\frac{u}{3}+\frac{5}{9}u^{2}+O(u^{3})\,. \label{RGGfu-sw-2L}%
\end{align}
The fixed point $u_{\ast}$ of the RGE determined by $\beta(u_{\ast})=0$ reads
\begin{equation}
u_{\ast}=\frac{\varepsilon}{3}+\frac{125}{486}\varepsilon^{2}+O(\varepsilon
^{3})\,. \label{FP-u}%
\end{equation}
At this fixed point, it follows that
\begin{equation}
\varepsilon - 2\kappa(u_{\ast}) = \gamma(u_{\ast}) = \eta\,. \label{ExRel}
\end{equation}
Hence, the static exponent $\eta$ and the dynamic exponent
\begin{equation}
z =2 +\zeta(u_{\ast}) \label{DynExp}
\end{equation}
are the only independent critical exponents of the problem.

Shift invariant observables free of redundancies are
\begin{equation}
M = \tau - g\langle s \rangle  \,, \quad H = 2gh + \tau^{2}\,, \label{Inv-MH}
\end{equation}
with RGEs
\begin{equation}
\mathcal{D}_{\mu}M = \kappa M\,, \quad \mathcal{D}_{\mu}H = (\gamma+\kappa)H\,.
\label{RG-MH}
\end{equation}
Note that $M$ and $H$ are linearly related to the Laplace transform $\Phi(z)$
of the cluster probability $\mathcal{P}(N)$ and the original Laplace variable
$z$, respectively. $M$ replaces the redundant parameter $\tau$ in all shift
invariant quantities. The integration of the RGEs (\ref{RG-MH}) at the fixed
point yields the equation of state \cite{BrJa81}
\begin{equation}
M = \text{const.}\, H^{\theta -2}\,. \label{EqSt}
\end{equation}

Likewise, by integrating the RGE~(\ref{RGG-sw}) at the fixed point,  we obtain
\begin{equation}
\langle s(\mathbf{r},t) \tilde{s}(\mathbf{0},0)\rangle = \frac{F(
t/r^z,rH^{\nu_{\text{P}}})}{r^{d-2+\eta+z}}\, \label{RespFu}
\end{equation}
for the response function. The static exponents are related by
\begin{equation}
\nu_{\text{P}} = \frac{2}{d-\eta}\,,\quad \theta -2 =
\frac{d-4+\eta}{d-\eta}\,,
\end{equation}
in conformity with the relation~(\ref{theta-nu}). For the dynamical exponent,
we obtain the 2-loop result
\begin{equation}
z=2+\zeta(u_{\ast})=2-\frac{\varepsilon}{9}-\frac{35}{18}\Big(\frac
{\varepsilon}{9}\Big)^{2}+O(\varepsilon^{3})\,. \label{z-sw-2L}%
\end{equation}

\begin{figure}[ptb]
\centering{\includegraphics[width=6.5cm]{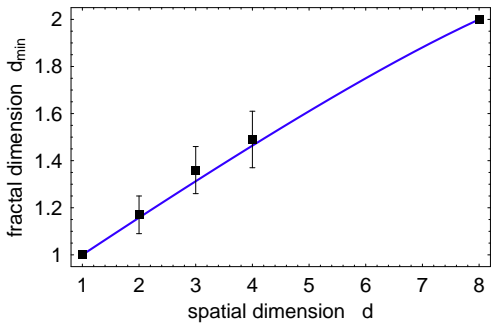}}%
\caption{(Color online) Rational-approximation estimate of the fractal
dimension of the minimal path
compared with numerical results of Ref.~\cite{HaDjMaStWe84}.}%
\label{fig:dmin}%
\end{figure}

\begin{figure}[ptb]
\centering{\includegraphics[width=6.5cm]{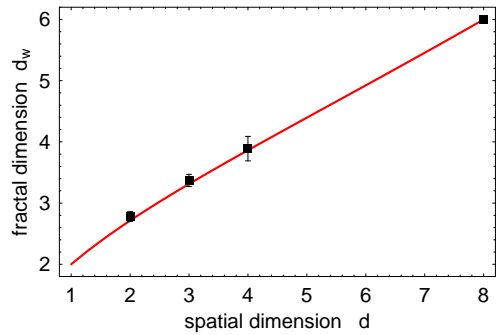}}%
\caption{(Color online) Rational-approximation estimate of the fractal
dimension of the random walk compared
with numerical results of Ref.~\cite{HaDjMaStWe84}.}%
\label{fig:dwalk}%
\end{figure}

Finally, we improve the numerical accuracy of this result by incorporating the
exact value $z=1$ for $d=1$ ($\varepsilon=7$) through the rational
approximation
\begin{equation}
d_{\min}=z\approx2-\varepsilon\frac{1134+245\varepsilon}{10206+1391\varepsilon
}\,. \label{dmin-rati}%
\end{equation}
Using relation (\ref{EinstRel}) together with $d_{f}=1/\nu_{\text{P}}$ and in
conjunction with the rational approximations given in Eqs.~ (\ref{nu-rati}) and
(\ref{dmin-rati}), we obtain the fractal dimensions displayed in
Table~\ref{tab:Fract-kritExp} and Figures~\ref{fig:dmin} and \ref{fig:dwalk}.
These figures also show the available simulation results~\cite{HaDjMaStWe84}.
We find that our field theoretical results agree remarkably well with the
latter.

\section{The collapsing RBP}
\label{sec:collapsingRBP}

Now, we turn to the collapse transition. As we did for the swollen phase, we
will first briefly review some known results~\cite{JaSt10} that we need as
input as we move along. The main task will be once again the calculation of the
dynamical exponent $z$ via the calculation of the dynamical renormalization
$Z_\lambda$.

\subsection{Static renormalizations and results}

As discussed in Sec.~\ref{sec:modelAndApproach}, $g_0$, $g_1$ and $g_2$ are all
relevant for the collapse transition and hence need to be kept in the field
theoretic functionals. As also discussed, $g_2^\prime$ is weakly irrelevant at
the collapse transition, and hence we can take the tree limit $g_2^\prime \to
0$ here. Thus, we work with the response functional $\mathcal{J}$ as given in
Eq.~(\ref{J-An}) and, as far as the static properties are concerned, the
quasi-static Hamiltonian $\mathcal{H}$ as given in Eq.~(\ref{H-An}).

Our renormalization scheme for the latter in dimensional regularization is%
\begin{subequations}
 \label{Ren-Sch-coll}%
\begin{align}
(\tilde{\varphi},\varphi,\bar{\psi},\psi)  &  \rightarrow(\mathring
{\tilde{\varphi}},\mathring{\varphi},\mathring{\bar{\psi}},\mathring{\psi
})=Z^{1/2}(\tilde{\varphi},\varphi+K\tilde{\varphi},\bar{\psi},\psi )\,,
\\
\underline{\tau}  &  \rightarrow\underline{\mathring{\tau}}=Z^{-1}%
\underline{\underline{Z}}\cdot\underline{\tau}\,,
\\
h  &  \rightarrow\mathring{h}=Z^{-1/2}(h+\frac{1}{2}G_{\varepsilon}^{1/2}%
\mu^{-\varepsilon/2}\underline{\tau}\cdot\underline{\underline{A}}%
\cdot\underline{\tau}){\,,}\
\\
G_{\varepsilon}^{1/2}g{_{\alpha}}  &  \rightarrow G_{\varepsilon}%
^{1/2}\mathring{g_{\alpha}}=Z^{-3/2}(u_{\alpha}+B_{\alpha})\mu^{\varepsilon
/2}\,,
\end{align}
\end{subequations}
where $\underline{\tau}=(\rho,\tau)$ and where $\varepsilon$ now measures the
deviation from $d=6$, $\varepsilon = 6-d$ . Since we are not interested here in
the renormalization of the control parameter $\rho$ that defines the cross-over
variable to the swollen phase and which goes to zero at the collapse- or
$\theta$-line, we set $\rho=0$ in the following calculations. Scaling invariant
combinations of the coupling constants are defined by%
\begin{equation}
v=u_{1}u_{2}\,,\qquad w=u_{0}u_{2}^{3}\,, \label{Def-vw}%
\end{equation}
with fixed point values%
\begin{subequations}
\label{vw-fix}%
\begin{align}
v_{\ast}  &  =0.6567\,(\varepsilon/6)+2.9707\,(\varepsilon/6)^{2}%
+O(\varepsilon^{3})\,,
\\
w_{\ast}  &  =0.7052\,(\varepsilon/6)^{2}+O(\varepsilon^{3})\,,
\end{align}
\end{subequations}
As it did for the swollen phase, the shift-invariance leads to Ward identities
for the collapse transition. Here, however, these Ward identities do not result
in a scaling relation between the polymer exponents $\theta$ and
$\nu_{\text{P}}$ but they nevertheless simplify the calculations or provide
consistency checks. The $\varepsilon$-expansions for these exponents are
\begin{subequations}
\label{PolExp-eps}%
\begin{align}
\theta &  =\frac{5}{2}-0.4925\,(\varepsilon/6)-0.5778\,(\varepsilon
/6)^{2}{\,,}
\\
\nu_{\text{P}}  &  =\frac{1}{4}+0.1915\,(\varepsilon/6)+0.0841\,(\varepsilon
/6)^{2}{\,.}
\end{align}
\end{subequations}

\subsection{Calculation of the dynamical scaling exponent of collapsing RBPs}

\begin{figure}[ptb]
\centering{\includegraphics[width=6cm]{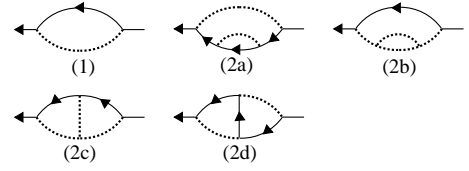}}\caption{Diagrams to
$2$-loop order for the collapse transition. Dotted lines symbolize all possible
arrangements of the static
propagator including ghosts.}%
\label{fig:dia-coll-2}%
\end{figure}

For our dynamical calculation, we complete the renormalization scheme
(\ref{Ren-Sch-coll}) by setting
\begin{subequations}
\label{Ren-Sch-coll-dyn}%
\begin{align}
s  &  \rightarrow\mathring{s}=Z^{1/2}(s+K\tilde{\varphi})\,,
\\
\tilde{s}  &  \rightarrow\mathring{\tilde{s}}=Z_{\lambda}Z^{1/2}\tilde
{s}\,,\qquad\lambda\rightarrow\mathring{\lambda}=Z_{\lambda}^{-1}\lambda\,.
\end{align}
\end{subequations}
Let us first describe the $1$-loop part of our calculation. The diagram (1) of
Fig.~(\ref{fig:dia-coll-2}) leads to the following contribution to the
selfenergy expanded to first order in $\omega$ and $\tau$%
\begin{align}
(1)_{c}  &  =-2\lambda g_{1}g_{2}\int_{\mathbf{p}}\frac{1}{\bigl(i\omega
/\lambda+\tau+\mathbf{p}^{2}\bigr)\bigl(\tau+(\mathbf{q}-\mathbf{p}%
)^{2}\bigr)}\nonumber\\
&  =-2\lambda g_{1}g_{2}\int_{\mathbf{p}}\Bigg\{\frac{1}{(\mathbf{p}%
-\mathbf{q})^{2}\mathbf{p}^{2}}-\frac{i\omega/\lambda+\tau}{(\mathbf{p}%
-\mathbf{q})^{2}\bigl(\mathbf{p}^{2}\bigr)^{2}}
\nonumber\\
&-\frac{\tau}{\mathbf{p}%
^{2}\bigl((\mathbf{p}-\mathbf{q})^{2}\bigr)^{2}}+\ldots\Bigg\}\,.
\label{Sc(1)}%
\end{align}
The factor $2$ stems from the two possible orientations of the propagator in
diagram (1). Performing the integrations as described in
Appendix~\ref{app:massless_calculationNearD=6} and carrying out an
 $\varepsilon$-expansion about $d=6$ we find the renormalized $1$-loop vertex function%
\begin{align}
\Gamma_{1,1}^{(1L)}(q,\omega)  &  =\bigl(ZZ_{\lambda}i\omega+Z_{\tau}%
\lambda\tau+Z\lambda q^{2}\bigr)\nonumber\\
&  -\frac{2v}{3\varepsilon}\bigl(3i\omega+6\lambda\tau+\lambda q^{2}%
\bigr)+\ldots\,. \label{Gammacoll-1L}%
\end{align}
Form this, we read off the $1$-loop renormalization factors%
\begin{equation}
Z=1+\frac{2v}{3\varepsilon}\,,\quad Z_{\tau}=1+\frac{4v}{\varepsilon}\,,\quad
Z_{\lambda}=1+\frac{4v}{3\varepsilon}\,, \label{Zcoll-1L}%
\end{equation}
with $Z$ and $Z_{\tau}$ as already known from \cite{JaSt10}.

Now we turn to the $2$-loop self-energy diagrams of Fig.~\ref{fig:dia-coll-2}.
For details of the calculation, we refer to
Appendix~\ref{app:massless_calculationNearD=6}. It encompasses the frame
counter-terms shown in Fig.~(\ref{fig:countcoll-2}) for which we get the
results compiled in Eq.~(\ref{Countcoll}). From these, we obtain the following
results for the counter-terms of the Feynman diagrams:
\begin{subequations}
\label{Countcoll-dia2L}
\begin{align}
\mathcal{C}(2a)_{c}   &=4\lambda(g_{1}g_{2})^{2} \Big\{Ct(3,1,1)
\nonumber\\
&-\bigl(2Ct(3,2,1) +3Ct(4,1,1)\bigr)\tau
\nonumber\\
&-\bigl(Ct(3,2,1)+2Ct(4,1,1)\bigr)\frac{i\omega}{\lambda }\Big\}\,,
\\
\mathcal{C}(2b)_{c}   &=\lambda\bigl(7(g_{1}g_{2})^{2}-g_{0}g_{2}{}
^{2}\bigr)\Big\{Ct(3,1,1)
\nonumber\\
&-\bigl(2Ct(3,2,1)+3Ct(4,1,1)\bigr)\tau-Ct(4,1,1)\frac{i\omega}{\lambda
}\Big\}\,,
\\
\mathcal{C}(2c)_{c}  & =\lambda\bigl(9(g_{1}g_{2})^{2}-g_{0}g_{2}{}
^{2}\bigr)\Big\{Ct(2,2,1)
\nonumber\\&-\bigl(4Ct(3,2,1)+Ct(2,2,2)\bigr)\tau-2Ct(3,2,1)\frac{i\omega}{\lambda
}\Big\}\,,
\\
\mathcal{C}(2d)_{c}   &=4\lambda(g_{1}g_{2})^{2}%
\Big\{Ct(2,2,1)
\nonumber\\
&-\bigl(4Ct(3,2,1)+Ct(2,2,2)\bigr)\tau\nonumber\\
& -\bigl(2Ct(3,2,1)+Ct(2,2,2)\bigr)\frac{i\omega}{\lambda }\Big\}\,.
\end{align}
\end{subequations}
These add up  to
\begin{align}
&\mathcal{C}\bigl((2a)+\cdots+(2d)\bigr)_{c}
\nonumber\\
&  =\Big[\Big(\frac{67}{18}%
-\frac{191}{216}\varepsilon\Big)v^{2}-\Big(\frac{5}{18}-\frac{13}%
{216}\varepsilon\Big)w\Big]\frac{\lambda q^{2}}{\varepsilon^{2}}\nonumber\\
&  +\Big[\Big(\frac{63}{2}-\frac{301}{24}\varepsilon\Big)v^{2}-\Big(\frac
{5}{2}-\frac{23}{24}\varepsilon\Big)w\Big]\frac{\lambda\tau}{\varepsilon^{2}%
}\nonumber\\
&  +\Big[\Big(\frac{25}{2}-\frac{103}{24}\varepsilon\Big)v^{2}-\Big(\frac
{5}{6}-\frac{11}{72}\varepsilon\Big)w\Big]\frac{i\omega}{\varepsilon^{2}%
}\,.\label{2_loop-countcoll}%
\end{align}
This defines the $2$-loop contributions to the renormalization factors $Z$,
$Z_{\tau}$, and $ZZ_{\lambda}$. Finally we obtain
\begin{subequations}
\label{Zcoll-2L}%
\begin{align}
Z &  =1+\frac{2v}{3\varepsilon}+\frac{1}{\varepsilon^{2}}\Big[\Big(\frac
{67}{18}-\frac{191}{216}\varepsilon\Big)v^{2}-\Big(\frac{5}{18}-\frac{13}%
{216}\varepsilon\Big)w\Big]\,
\\
Z_{\tau} &  =1+\frac{4v}{\varepsilon}+\frac{1}{\varepsilon^{2}}\Big[\Big(\frac
{63}{2}-\frac{301}{24}\varepsilon\Big)v^{2}-\Big(\frac{5}{2}-\frac{23}%
{24}\varepsilon\Big)w\Big]\, ,
\\
Z_{\lambda} &  =1+\frac{4v}{3\varepsilon}+\frac{1}{\varepsilon^{2}%
}\Big[\Big(\frac{83}{9}-\frac{92}{27}\varepsilon\Big)v^{2}-\Big(\frac{5}%
{9}-\frac{5}{24}\varepsilon\Big)w\Big]\,.
\end{align}
\end{subequations}
As they should be, the first two are in conformity with results of
\cite{JaSt10}. The dynamic renormalization group function becomes
\begin{equation}
\zeta=\left.  \frac{\partial\ln Z_{\lambda}}{\partial\ln\mu}\right\vert
_{0}=-\frac{4v}{3}+\Big(\frac{184}{27}v^{2}-\frac{5}{12}w\Big)+O(3\text{-loop}%
)\,.\label{zeta-coll}%
\end{equation}
Using the fixed point values of the coupling constant given in
Eq.~(\ref{vw-fix}), we obtain the $\varepsilon$-expansion of the dynamic
exponent and by the same token the scaling dimension of the minimal path:
\begin{equation}
d_{\min}=z=2+\zeta_{\ast}=2-0.8756\,(\varepsilon/6)-1.3162\,(\varepsilon
/6)^{2}+O(\varepsilon^{3})\,.\label{dmin-scoll}%
\end{equation}
Taking into account the exact values $z=\nu_{\text{P}}=1$ for $d=1$
($\varepsilon =5$), we propose the rational approximations
\begin{equation}
d_{\min}=z\approx2-0.14593\,\varepsilon\frac{1+0.726717\,\varepsilon
}{1+0.476179\,\varepsilon}\,,\label{zeta-coll-appr}%
\end{equation}
and
\begin{equation}
1/d_{f}=\nu_{\text{P}}\approx\frac{1}{4}+\frac{0.0319167\,\varepsilon}%
{1-0.0733194\,\varepsilon-0.016825\,\varepsilon^{2}}\,, \label{nu-coll-appr}%
\end{equation}
the latter being based on the $\varepsilon$-expansion of $\nu_{\text{P}}$ as
given in Eq.~(\ref{PolExp-eps}). Table~\ref{tab:Fractcoll-kritExp} lists the
numerical values resulting from these rational approximations for various
spatial dimension.
\begin{table}[ptb]%
\caption{Estimates of the fractal dimensions of collapsing RBPs obtained via
rational approximations.}%
\begin{tabular}
[c]{c|cccccc}\hline\hline $d$ & \quad$1$ & \quad$2$ & \quad$3$ & \quad$4$ &
\quad$5$ & \quad$6$\\\hline $d_{w}$ & \quad$2$ & \quad$3.038$ & \quad$3.882$ &
\quad$4.631$ & \quad$5.332$
& \quad$6$\\
$d_{\min}$ & \quad$1$ & \quad$1.192$ & \quad$1.396$ & \quad$1.612$ &
\quad$1.824$ & \quad$2$\\
$d_{f}$ & \quad$1$ & \quad$1.846$ & \quad$2.486$ & \quad$3.019$ & \quad$3.508$
& \quad$4$\\\hline\hline
\end{tabular}
\label{tab:Fractcoll-kritExp}%
\end{table}

\section{Concluding remarks}
\label{sec:concludingRemarks} In summary, we have studied diffusion and
transport on swollen and collapsing randomly polymers using renormalized
dynamical field theory. In particular, we have calculated the diffusion
exponent and a set of transport exponents including the fractal dimension of
the minimal path and the resistance exponent to 2-loop order. For the swollen
polymer, our results are an improvement of the previously known 1-loop results.
For the collapse transition, our results are entirely new in the sense that
hitherto no results beyond mean-field theory existed.

From a conceptual or diagrammatic standpoint, it was interesting to see that
the dynamical Feynman diagrams for the self-energy decompose into dynamic part
which has the form of a SAW connecting the external legs and a residual
quasi-static part. This observation simplified the calculation of the dynamical
renormalization enough to enable us to treat the problem to 2-loop order.

As far as we know, the existing numerical results for diffusion and transport
on substrates in the universality class randomly branched polymers and lattice
animals are more than 25 years old. For the collapse transition, no such
results exist at all. We hope that our results encourage simulation work
leading to improved and extended numerical results.

\begin{acknowledgments}
This work was supported in part (OS) by NSF-DMR-1104707.
\end{acknowledgments}

\appendix

\section{$2$-Loop massless calculation of frames and counter-terms near $d=8$ }
\label{app:massless_calculation}

\begin{figure}[ptb]
\centering{\includegraphics[width=6.5cm]{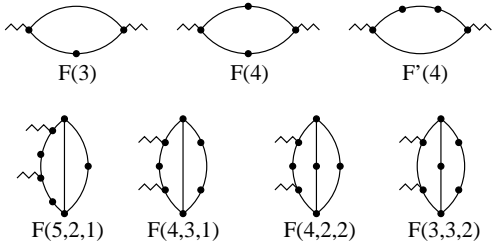}}\caption{Frames of the
diagrams for swollen RBPs up to 2-loop order.}%
\label{fig:frames-mit}%
\end{figure}

Here, we calculate the frames of the dynamical diagrams for the swollen phase
using dimensional regularization \cite{tHoVe72/73}. We consider only those
frames that are required to determine the dynamic renormalization factor to
$2$-loop order. The calculation can be greatly simplified via the consequent
use of massless propagators \cite{Ka83/85}. Of course, massless propagators
bring about the danger of infrared (IR) singularities which would require the
introduction of IR-counter-terms \cite{ChTk82,ChSm84} if they indeed occurred.
However, cleverly using infrared rearrangements of external momenta
\cite{Vl78/79/80,ChKaTk80/81}, the IR-singularities can be avoided at least up
to $3$-loop order. Figure~\ref{fig:frames-mit} lists the massless frames that
we need for our 2-loop calculation. The wiggly lines symbolize external momenta
$\pm \mathbf{q}$ flowing into or out of the frame. These wiggly lines are
positioned so that the frames are free of IR-singularities.

First, let us return to the 1-loop part of our calculation sketched in
Sec.~\ref{sec:swollenRBP}. Equation~(\ref{S(1-2)}) for diagram (1) can be
expressed as
\begin{align}
(1)  &  =2\lambda g^{2}\Big\{-G_{d}(1,2)+G_{d}(2,2)(i\omega/\lambda+\tau )
\nonumber\\
&+G_{d}(1,3)2\tau+\ldots\Big\}\,, \label{S(1-3)}%
\end{align}
where $G_{d}(\cdots)$ is defined through the fundamental integral
\begin{equation}
\int_{\mathbf{p}}\frac{1}{\bigl(\mathbf{p}^{2}\bigr)^{\lambda_{1}%
}\bigl((\mathbf{p}-\mathbf{q})^{2}\bigr)^{\lambda_{2}}}=G_{d}(\lambda
_{1},\lambda_{2})q^{d-2(\lambda_{1}+\lambda_{2})}\,. \label{G-d-def}%
\end{equation}
This integral can be easily carried out using standard methods with the result
\begin{equation}
G_{d}(\lambda_{1},\lambda_{2})=\frac{\Gamma(d/2-\lambda_{1})\Gamma
(d/2-\lambda_{2})\Gamma(\lambda_{1}+\lambda_{2}-d/2)}{(4\pi)^{d/2}%
\Gamma(\lambda_{1})\Gamma(\lambda_{2})\Gamma(d-\lambda_{1}-\lambda_{2})}\,.
\label{G-d}%
\end{equation}
$\varepsilon$-expansion in $\varepsilon=8-d$ leads to
\begin{subequations}
 \label{list-G-d-1}%
\begin{align}
F(3)  &  =G_{d}(1,2)q^{2-\varepsilon}=-q^{2}\frac{(\mu/q)^{\varepsilon}}
{6\varepsilon}\Big(1+\frac{4}{3}\varepsilon+\ldots\Big)
G_{\varepsilon}\mu^{-\varepsilon},
\\
F(4)  &  =G_{d}(1,3)q^{-\varepsilon}=\frac{(\mu/q)^{\varepsilon}}
{3\varepsilon}\Bigl(1+\frac{13}{12}\varepsilon+\ldots
)\Bigr)G_{\varepsilon}\mu^{-\varepsilon}\,,
\\
F^{\prime}(4)  &  =G_{d}(2,2)q^{-\varepsilon}=\frac{(\mu/q)^{\varepsilon}%
}{3\varepsilon}\Bigl(1+\frac{5}{6}\varepsilon+\ldots
)\Bigr)G_{\varepsilon}\mu^{-\varepsilon}\,,
\end{align}
\end{subequations}
with $G_{\varepsilon}=\Gamma(1+\varepsilon/2)/(4\pi)^{d/2}$. Note that for
these values of the parameters $\lambda_{i}$, the integral (\ref{G-d-def}) does
not contain IR-singularities. Hence, the $\varepsilon $-poles in
Eqs.~(\ref{list-G-d-1}) purely arise from UV-singularities as they should,
i.e., we have successfully avoided  IR-singularities even though we resorted to
massless propagators.

Next, we consider the $2$-loop diagrams. The frames with the arrangement of
external momenta as shown in Fig.~(\ref{fig:frames-mit}) are free of
IR-singularities. They are easily integrated through a successive application
of Eq.~(\ref{G-d-def}). We obtain
\begin{subequations}
\label{list-G-d-2}%
\begin{align}
F(5,2,1)  &  =G_{d}(2,6-d/2)G_{d}(1,2)q^{-2\varepsilon}\nonumber\\
&  =-\frac{(\mu/q)^{2\varepsilon}}{36\varepsilon^{2}}\Bigl(1+\frac{25}%
{12}\varepsilon+\ldots\Bigr)G_{\varepsilon}^{2}\mu^{-2\varepsilon }\,,
\\
F(4,3,1)  &  =G_{d}(2,6-d/2)G_{d}(1,3)q^{-2\varepsilon}\nonumber\\
&  =\frac{(\mu/q)^{2\varepsilon}}{18\varepsilon^{2}}\Bigl(1+\frac{11}%
{6}\varepsilon+\ldots\Bigr)G_{\varepsilon}^{2}\mu^{-2\varepsilon }\,,
\\
F(4,2,2)  &  =G_{d}(2,6-d/2)G_{d}(2,2)q^{-2\varepsilon}\nonumber\\
&  =\frac{(\mu/q)^{2\varepsilon}}{18\varepsilon^{2}}\Bigl(1+\frac{19}%
{12}\varepsilon+\ldots\Bigr)G_{\varepsilon}^{2}\mu^{-2\varepsilon }\,,
\\
F(3,3,2)  &  =G_{d}(1,7-d/2)G_{d}(2,3)q^{-2\varepsilon}\nonumber\\
& =\frac{(\mu/q)^{2\varepsilon}}{24\varepsilon}\Bigl(1+\ldots\Bigr)
G_{\varepsilon}^{2}\mu^{-2\varepsilon}\,.
\end{align}
\end{subequations}

Now, we calculate the counter-terms using a BPHZ-type of renormalization
\cite{ItZu80,Vl78/79/80}, i.e., a recursive renormalization procedure defined
by a sequence of operative steps. To this end, we define a pole-separating
procedure
$\mathfrak{P}$ operating on $\varepsilon$-expansions%
\begin{equation}
\mathfrak{P}\sum_{i=-\infty}^{\infty}c_{i}\varepsilon^{i}=\sum_{i=-\infty
}^{-1}c_{i}\varepsilon^{i}\,. \label{PolSep}%
\end{equation}
The counter-terms are constructed by the operation
\begin{equation}
\mathfrak{C}=\mathfrak{PR}^{\prime}\,, \label{CountDef}%
\end{equation}
where the incomplete renormalization $\mathfrak{R}^{\prime}$ operates on the
momentum-integrals $I_{\Gamma}$ of the one-particle irreducible
diagrams $\Gamma$. It is recursively defined by%
\begin{equation}
\mathfrak{R}^{\prime}I_{\Gamma}=I_{\Gamma}+\sum_{\{\gamma\}\in D_{\Gamma}%
}I_{\Gamma/\{\gamma\}}\cdot\prod_{\gamma_{i}\in\{\gamma\}}(-\mathfrak{C}%
I_{\gamma_{i}})\,. \label{IncRen}%
\end{equation}
Here $D_{\Gamma}$ is the set of all collections $\{\gamma\}=\{\gamma
_{1},\gamma_{2},\ldots\}$ of disjunct superficially divergent one-particle
irreducible subdiagrams $\gamma_{i}$ of $\Gamma$, and $\Gamma/\{\gamma\}$ is
the diagram obtained from $\Gamma$ by collapsing each of these subdiagrams to
points. Applying $\mathfrak{R}^{\prime}$ to a superficially convergent diagram
$\Gamma$, $\mathfrak{R}^{\prime}I_{\Gamma}$ is finite for $\varepsilon
\rightarrow0$, and $\mathfrak{C}I_{\Gamma}=0$. The operation $\mathfrak{R}%
^{\prime}$ applied to superficially divergent diagrams produces poles in
$\varepsilon$ with coefficients polynomial in the external momenta and internal
masses without non-primitive logarithmic terms. Finally, the operation
$\mathfrak{C}I_{\Gamma}$ yields the wanted counter-terms, and the complete
renormalization $\mathfrak{R}I_{\Gamma}=\mathfrak{R}^{\prime}I_{\Gamma
}-\mathfrak{C}I_{\Gamma}=(1-\mathfrak{P})\mathfrak{R}^{\prime}I_{\Gamma}$ leads
to a finite result.

\begin{figure}[ptb]
\centering{\includegraphics[width=5cm]{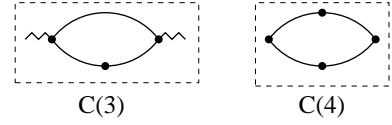}}\caption{1-loop
counter-terms in $d=8-\varepsilon$ dimensions.}%
\label{fig:count-1}%
\end{figure}
Of course, applying $\mathfrak{R}^{\prime}$ to $1$-loop diagrams is trivial.
The counter-terms (dashed boxes) are shown in Fig.~\ref{fig:count-1}. Using
Eqs.~(\ref{list-G-d-1}), we get%
\begin{subequations}
\label{Count-1loop}%
\begin{align}
C(3)  &  =\mathfrak{P}F(3)=-\frac{G_{\varepsilon}\mu^{-\varepsilon}%
}{6\varepsilon}q^{2}\,,
\\
C(4)  &  =\mathfrak{P}F(4)=\mathfrak{P}F^{\prime}(4)=\frac{G_{\varepsilon}%
\mu^{-\varepsilon}}{3\varepsilon}\,.
\end{align}
\end{subequations}
Note that we leave factors $G_{\varepsilon}\mu^{-\varepsilon}$ un-expanded
because they are absorbed by the dimension-bearing coupling constants, cf.\ the
renormalization scheme (\ref{Ren-Sch}). Note also that because of the
application of the operation $\mathfrak{C}$, it does not matter at which
vertices we have injected the external momenta as IR-regulators into a given
superficially logarithmically divergent diagram $\Gamma$ as long as
IR-divergencies are avoided.

\begin{figure}[ptb]
\centering{\includegraphics[width=6cm]{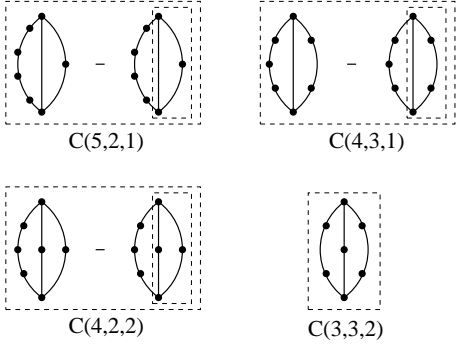}}\caption{\label{fig:count-2}
2-loop counter-terms in $d=8-\varepsilon$ dimensions.}
\end{figure}

At $2$-loop order, the procedure $\mathfrak{P}$ produces the counter-terms
listed in Fig.~\ref{fig:count-2}. Mathematically, these stand for
\begin{subequations}
\begin{align}
&  C(5,2,1)=\mathfrak{P}\bigl[F(5,2,1)-G_{d}(2,2)C(3)\bigr]\nonumber\\
&  =\mathfrak{P}\Big\{-\frac{(\mu/q)^{2\varepsilon}}{36\varepsilon^{2}%
}\Big(1+\frac{25}{12}\varepsilon\Big)+\frac{(\mu/q)^{\varepsilon}%
}{18\varepsilon^{2}}\Big(1+\frac{5}{6}\varepsilon
\Big)\Big\}\bigl(G_{\varepsilon}\mu^{-\varepsilon}\bigr)^{2}\nonumber\\
&  =\frac{\bigl(G_{\varepsilon}\mu^{-\varepsilon}\bigr)^{2}}{36\varepsilon
^{2}}\Big(1-\frac{5}{12}\varepsilon\Big)\,, \label{Count-2loop-a}%
\end{align}%
\begin{align}
&  C(4,3,1)=\mathfrak{P}\bigl[F(4,3,1)-G_{d}(2,2)C(4)\bigr]\nonumber\\
&  =\mathfrak{P}\Big\{\frac{(\mu/q)^{2\varepsilon}}{18\varepsilon^{2}%
}\Big(1+\frac{11}{6}\varepsilon\Big)-\frac{(\mu/q)^{\varepsilon}}%
{9\varepsilon^{2}}\Big(1+\frac{5}{6}\varepsilon\Big)\Big\}\bigl(G_{\varepsilon
}\mu^{-\varepsilon}\bigr)^{2}\nonumber\\
&  =-\frac{\bigl(G_{\varepsilon}\mu^{-\varepsilon}\bigr)^{2}}{18\varepsilon
^{2}}\Big(1-\frac{1}{6}\varepsilon\Big)\,, \label{Count-2loop-b}%
\end{align}%
\begin{align}
&  C(4,2,2)=\mathfrak{P}\bigl[F(4,2,2)-G_{d}(2,2)C(4)\bigr]\nonumber\\
&  =\mathfrak{P}\Big\{\frac{(\mu/q)^{2\varepsilon}}{18\varepsilon^{2}%
}\Big(1+\frac{19}{12}\varepsilon\Big)-\frac{(\mu/q)^{\varepsilon}%
}{9\varepsilon^{2}}\Big(1+\frac{5}{6}\varepsilon
\Big)\Big\}\bigl(G_{\varepsilon}\mu^{-\varepsilon}\bigr)^{2}\nonumber\\
&  =-\frac{\bigl(G_{\varepsilon}\mu^{-\varepsilon}\bigr)^{2}}{18\varepsilon
^{2}}\Big(1+\frac{1}{12}\varepsilon\Big)\,, \label{Count-2loop-c}%
\end{align}%
\begin{equation}
C(3,3,2)=\mathfrak{P}\bigl[F(3,3,2)\bigr]=\frac{\bigl(G_{\varepsilon}%
\mu^{-\varepsilon}\bigr)^{2}}{24\varepsilon}\,. \label{Count-2loop-d}%
\end{equation}
\end{subequations}

\section{$2$-Loop calculation and counter-terms near $d=6$}
\label{app:massless_calculationNearD=6}

\begin{figure}[ptb]
\centering{\includegraphics[width=5cm]{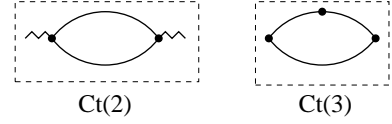}}\caption{1-loop
counter-terms in $d=6-\varepsilon$ dimensions.} \label{fig:countcoll-1}
\end{figure}

\begin{figure}[ptb]
\centering{\includegraphics[width=6.5cm]{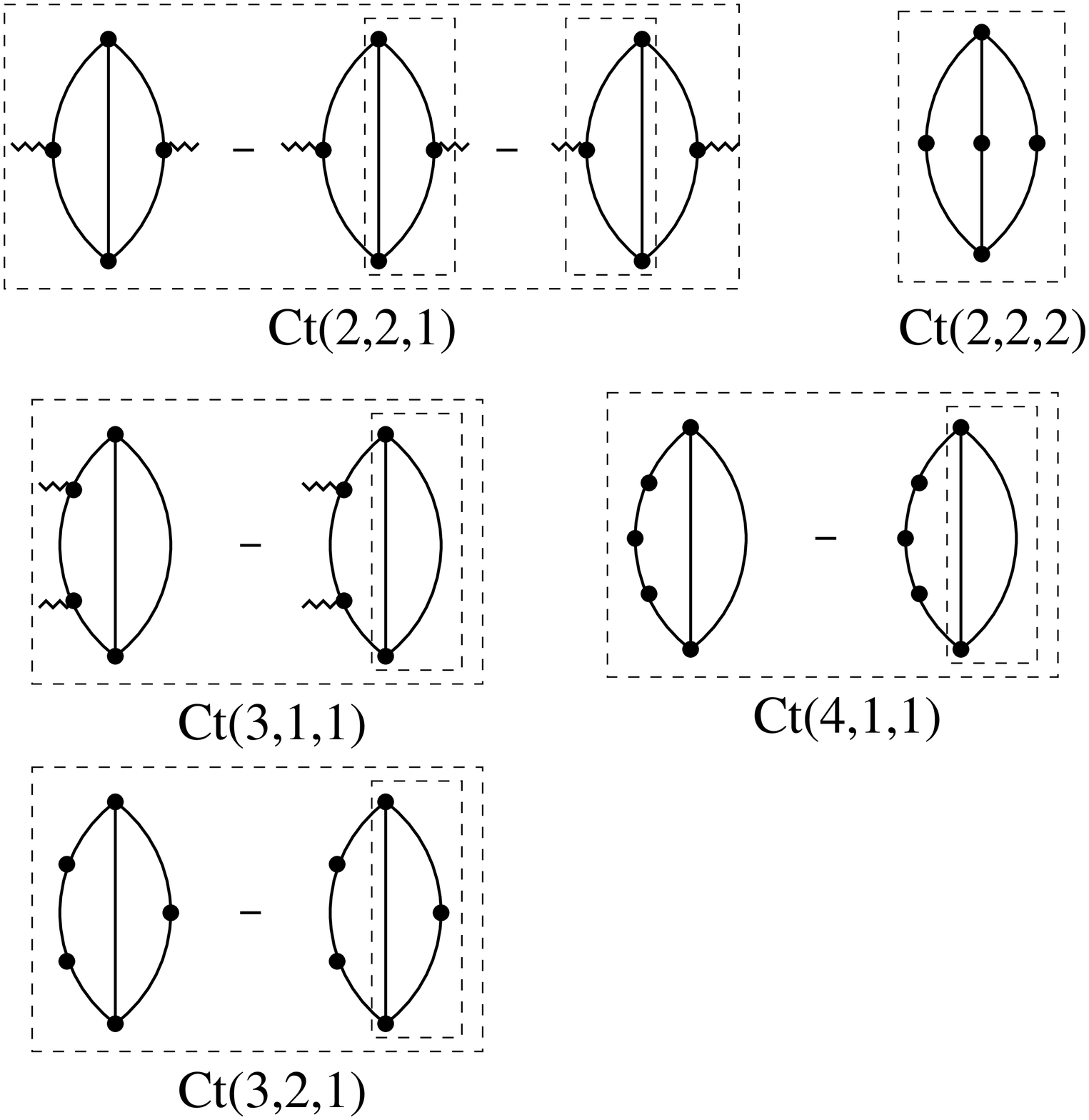}}\caption{2-loop
counter-terms in $d=6-\varepsilon$ dimensions.}%
\label{fig:countcoll-2}%
\end{figure}

The counter-terms needed for the collapse transition, see
Figs.~\ref{fig:countcoll-1} and \ref{fig:countcoll-2}, can be obtained through
a massless calculation using methods similar to those explained in
Appendix~\ref{app:massless_calculation} in conjunction with
 t'Hooft and Veltman's `partial $p$' method \cite{tHoVe72/73}. Partially, they can also be extracted from results of de
Alcantara Bonfim \textit{et al}.~\cite{AKM81} for the usual $\phi^{3}$-field
theory.

Let us start here by revisiting Eq.~(\ref{Sc(1)}). In terms of the fundamental
integral~(\ref{G-d-def}), this 1-loop diagram can be expressed as
\begin{align}
(1)_{c}  &    =2\lambda g_{1}g_{2}\Big\{-G_{d}(1,1)+G_{d}(2,1)(i\omega/\lambda
+2\tau)+\ldots\Big\}\,. \label{Sc(1)InTermsOfG}%
\end{align}
After $\varepsilon$-expansion about $d=6$, we extract the counter-terms
\begin{align}
&  Ct(2)=-\frac{G_{\varepsilon}\mu^{-\varepsilon}}{3\varepsilon}q^{2}\,,\quad
Ct(3)=\frac{G_{\varepsilon}\mu^{-\varepsilon}}{\varepsilon}\,,\
\end{align}

For the 2-loop counter-terms, we obtain
\begin{subequations}
\label{Countcoll}%
\begin{align}
&  Ct(2,2,1)=\frac{\bigl(G_{\varepsilon}\mu^{-\varepsilon}\bigr)^{2}%
}{3\varepsilon^{2}}\Big(1-\frac{1}{3}\varepsilon\Big)q^{2}\,,
\\
&  Ct(3,1,1)=-\frac{\bigl(G_{\varepsilon}\mu^{-\varepsilon}\bigr)^{2}%
}{18\varepsilon^{2}}\Big(1-\frac{11}{12}\varepsilon\Big)q^{2}\,,
\\
&  Ct(4,1,1)=\frac{\bigl(G_{\varepsilon}\mu^{-\varepsilon}\bigr)^{2}%
}{6\varepsilon^{2}}\Big(1-\frac{7}{12}\varepsilon\Big)\,,
\\
&  Ct(3,2,1)=-\frac{\bigl(G_{\varepsilon}\mu^{-\varepsilon}\bigr)^{2}%
}{2\varepsilon^{2}}\Big(1-\frac{1}{4}\varepsilon\Big)\,,
\\
&  Ct(2,2,2)=\frac{\bigl(G_{\varepsilon}\mu^{-\varepsilon}\bigr)^{2}%
}{2\varepsilon}\,.
\end{align}
\end{subequations}

\section{Relations between Feynman integrals in $d$ and $(d-2)$ dimensions}
\label{app:RelationsBetweenFeynmanIntegrals}

It is well known that supersymmetry relates many Feynman integrals with
insertions to the corresponding integrals in $2$ lesser dimensions
\cite{PaSo79,PaSo81}. Therefore, it arises the question whether such relations
can be used to determine the counter-terms of the frames needed in our
dynamical calculation.  We will show explicitly that this indeed is the case at
least to 2-loop order. This observation may open the route to a $3$-loop
calculation since all counterterms near $d=6$ are known \cite{AKM81}. For
background information on the following reasoning, see the textbook by Itzykson
and Zuber \cite{ItZu80}.

Consider a $1$-particle irreducible Feynman diagram $G$ without tadpoles
consisting of $V$ vertices, $I$ internal lines, and $L=I-V+1$ loops in $d$
dimensions. Each line $l$ carries a propagator $1/(\tau
_{l}+\mathbf{q}_{l}^{\,2})$ where $\mathbf{q}_{l}$ is a $d$-dimensional
momentum vector, and $\tau_{l}$ an auxiliary mass squared which is finally set
to zero in a massless calculation. Next, let us introduce an (arbitrary)
orientation of each line, and we define the incidence matrix
$(\varepsilon_{vl})$,
\begin{equation}
\varepsilon_{vl}=\left\{
\begin{array}
[c]{rl}
1 & \text{if the vertex }v\text{ is the starting point of line }l\\
-1 & \text{if the vertex }v\text{ is the endpoint of line }l\\
0 & \text{if }l\text{ is not incident on }v\,.
\end{array}
\right.  \label{InzMat}
\end{equation}
Momentum conservation at each vertex $v$ is expressed as
\begin{equation}
\sum_{l=1}^{I}\varepsilon_{vl}\mathbf{q}_{l}=\mathbf{Q}_{v}\,, \label{MomCons}
\end{equation}
where $\mathbf{Q}_{v}$ is the external momentum flowing into vertex $v$. Only
$(V-1)$ of the conservation laws (\ref{MomCons}) are independent since
\begin{equation}
\sum_{v=1}^{V}\varepsilon_{vl}=0\,. \label{InzId}
\end{equation}
The last one produces the overall conservation
\begin{equation}
\sum_{v=1}^{V}\mathbf{Q}_{v}=0\,. \label{OvCont}
\end{equation}

The contribution $I(G,\{\mathbf{Q}_{v}\})_{d}$ of the diagram $G$ to the
corresponding vertex function, can be written in a Schwinger parametric form as
\cite{ItZu80}
\begin{align}
I(G,\{\mathbf{Q}_{v}\})_{d}  &  =\int_{0}^{\infty}\prod\limits_{l=1}^{I}
ds_{l}\int_{\{\mathbf{q}\}}\exp\Big(-\sum_{l}\bigl(\tau_{l}+\mathbf{q}_{l}
^{2}\bigr)s_{l}\Big)\nonumber\\
&  \times\prod\limits_{v=1}^{V-1}\Big((2\pi)^{d}\delta^{(d)}\bigl(\sum
_{l}\varepsilon_{vl}\mathbf{q}_{l}-\mathbf{Q}_{v}\bigr)\Big)\,,
\label{FeynmInt}
\end{align}
where we focus on the diagram's frame (integral over the loop-momenta) and omit
any symmetry factors and coupling constants. Using the integral representation
of the $\delta$-functions and performing all the arising Gaussian integration
we arrive at
\begin{align}
I(G,\{\mathbf{Q}_{v}\})_{d}&=\int_{0}^{\infty}\prod\limits_{l=1}^{I}
\bigl(\mathrm{e}^{-\tau_{l}s_{l}}ds_{l}\bigr)\nonumber\\
&\times\frac{\exp\Big(-\mathcal{Q}
(\{s_{l}\},\{\mathbf{Q}_{v}\})/\mathcal{P}(\{s_{l}\})\Big)}{\bigl((4\pi
)^{L}\mathcal{P}(\{s_{l}\})\bigr)^{d/2}}\,. \label{FeynmPar}
\end{align}
Here, $\mathcal{P}(\{s_{l}\})$ is given by
\begin{equation}
\mathcal{P}(\{s_{l}\})=(4\pi)^{-L}\prod\limits_{l=1}^{I}(4\pi s_{l}
)\,\frac{\det A}{(4\pi)^{V-1}}\,. \label{Ps}
\end{equation}
$A$ is the $(V-1)\times(V-1)$ matrix with elements $A_{vv^{\prime}
}=\sum_{l}\varepsilon_{vl}s_{l}^{-1}\varepsilon_{v^{\prime}l}$ where
$v,v^{\prime}=1,\ldots,V-1$. $\mathcal{Q}(\{s_{l}\},\{\mathbf{Q}_{v}\})$ is the
bilinear form constructed from the $\{\mathbf{Q}_{v}\}$ and the inverse of the
matrix $A$:
\begin{equation}
\mathcal{Q}(\{s_{l}\},\{\mathbf{Q}_{v}\})=\mathcal{P}(\{s_{l}\})\sum
_{v,v^{\prime}=1}^{V-1}\mathbf{Q}_{v}\cdot (A^{-1})_{vv^{\prime}}
\mathbf{Q}_{v^{\prime}}\,. \label{Qs}%
\end{equation}
$\mathcal{P}$ and $\mathcal{Q}$ are readily simplified using Kirchhoff's laws
on linear electrical networks \cite{Kirch47},
\begin{subequations}
\label{Theorems}
\begin{align}
\mathcal{P}(\{s_{l}\})  &  =\sum_{T}\prod\limits_{l\notin T}s_{l}
\,,\label{TheoremP}\\
\mathcal{Q}(\{s_{l}\},\{\mathbf{Q}_{v}\})  &  =\sum_{(T_{1},T_{2})}
\prod\limits_{l\notin T_{1},T_{2}}s_{l}\bigl(\sum_{v\in T_{1}}\mathbf{Q}
_{v}\bigr)^{2}\,. \label{TheoremQ}
\end{align}
\end{subequations}
Here, the sum $T$ runs over all spanning trees, and the sum $(T_{1},T_{2})$
runs over all pairs of mutually disconnecteds panning trees (a forest with two
trees) on the diagram $G$, respectively. Note that the dependence of
$I(G,\{\mathbf{Q}_{v}\})_{d}$ as given in Eq.\thinspace(\ref{FeynmInt}) on the
dimensionality $d$ entirely rests in the exponent of the denominator.

Now, let us define the procedure
\begin{align}
\mathcal{T}^{\ast}=(4\pi)^{L}\mathcal{P} (\{-\partial/\partial\tau_{l}\})
\end{align}
that produces the sum over all diagrams obtained by inserting $s^2$ into all
lines of $G$ not belonging to any tree. Applying this procedure to
$I(G,\{\mathbf{Q}_{v}\})_{d}$ transforms this integral into its
$(d-2)$-dimensional counterpart:
\begin{align}
\mathcal{T}^{\ast}I(G,\{\mathbf{Q}_{v}\})_{d}&=I(\mathcal{T}^{\ast
}G,\{\mathbf{Q}_{v}\})_{d}\nonumber\\
&=I(G,\{\mathbf{Q}_{v}\})_{d-2}\,. \label{TheoremT}
\end{align}

Next, we turn to the self-energy diagrams in $d$ dimensions. These diagrams
have only 2 external momenta,  $\{\mathbf{Q}_{v}\}=\{\mathbf{q}$,
$-\mathbf{q\}}$, and the bilinear form
$\mathcal{Q}(\{s_{l}\},\{\mathbf{Q}_{v}\})$ can contain only those partitions
into two disconnected trees $(T_{1},T_{2})$ that disconnect the two external
vertices where the external momenta enter or exit the diagram. The product
$\prod \limits_{l\notin T_{1},T_{2}}s_{l}$ runs therefore over the $L+1$ lines
that belong to a cut-set which divides the diagram into two tree-like
subdiagrams each containing one external vertex. In this case, the form
$\mathcal{Q} (\{s_{l}\},\{\mathbf{Q}_{v}\})$ reduces to
\begin{equation}
\mathcal{Q}(\{s_{l}\},\{\mathbf{Q}_{v}\})=\sum_{(T_{1},T_{2})}\prod
\limits_{l\notin T_{1},T_{2}}s_{l}\,\mathbf{q}^{2}=:\mathcal{Q}^{\ast}
(\{s_{l}\})\mathbf{q}^{2}\,, \label{DefCst}%
\end{equation}
where $v_{1}\in T_{1}$ and $v_{2}\in T_{2}$. Now, we define the procedure
\begin{equation}
\mathcal{C}^{\ast}
=(4\pi)^{L}\mathcal{Q}^{\ast}(\{-\partial/\partial\tau_{l}\})
\end{equation}
which produces the sum over all diagrams obtained from $G$ by placing an
insertion into any of its cut-set lines. Applying this procedure to
$I(G,\{\mathbf{q},-\mathbf{q\}})_{d}$ results in
\begin{align}
\mathcal{C}^{\ast}I(G,\{\mathbf{q},-\mathbf{q\}})_{d}&=I(\mathcal{C}^{\ast
}G,\{\mathbf{q},-\mathbf{q\}})_{d}\nonumber\\
&=-\frac{\partial}{\partial\mathbf{q}^{2}
}I(G,\{\mathbf{q},-\mathbf{q\}})_{d-2}\,. \label{TheoremC}
\end{align}
Note that this theorem relates quadratically diverging diagrams to logarithmic
diverging diagrams in 2 dimensions higher.

Having derived the theorems (\ref{TheoremT}) and (\ref{TheoremC}), we now can
use them to extract relations between the counter-terms encountered in our
2-loop calculations. Applying the $\mathcal{T}^{\ast}$-procedure, we obtain
\begin{align}
Ct(2,2,2)  &  =12C(3,3,2))\,,\nonumber\\
Ct(3,2,1)  &  =6C(4,3,1)+3C(4,2,2)+2C(3,3,2)\,,\nonumber\\
Ct(4,1,1)  &  =8C(5,2,1)+C(4,2,2\,. \label{RelCount-T}
\end{align}
The $\mathcal{C}^{\ast}$-procedure provides us with the relations
\begin{align}
-\frac{\partial}{\partial\mathbf{q}^{2}}Ct(2,2,1)  &
=4C(4,3,1)+2C(4,2,2)+2C(3,3,2)\,,\nonumber\\
-\frac{\partial}{\partial\mathbf{q}^{2}}Ct(3,1,1)  & =4C(5,2,1)+C(4,2,2)\,.
\label{RelCount-C}
\end{align}
These five relations determine all the four 2-loop counter-terms near $8$
dimensions from the five 2-loop counter-terms near $6$ dimensions and in
addition to that provide us one supplementary consistency check for our
calculations.

\end{document}